%% file: manuscript.tex
\documentclass[lettersize,journal]{IEEEtran} 
\usepackage{amsmath,amsfonts,amssymb}
\usepackage{dsfont}
\usepackage{algorithmic}
\usepackage{algorithm}
\usepackage{array}
\usepackage[caption=false,font=footnotesize,labelfont={bf,rm},textfont=rm]{subfig}
\usepackage{textcomp}
\usepackage{stfloats}
\usepackage{cite}
\usepackage{url}
\usepackage{color}
\usepackage{xcolor}
\usepackage{colortbl}
\usepackage{soul}

\usepackage{verbatim}
\usepackage{graphicx}
\usepackage{subcaption} 
\usepackage{epsfig}
\usepackage{ulem}
\usepackage{tikz}
\usetikzlibrary{decorations.pathreplacing,arrows.meta}
\usepackage{booktabs}
\usepackage{float}
\usepackage{epstopdf}
\usepackage{cleveref}
\usepackage{amsthm}
\newtheoremstyle{mystyle}
  {6pt}    
  {6pt}    
  {\normalfont}  
  {0pt}    
  {\itshape}    
  {:}      
  {5pt}    
  {\thmname{#1}\thmnumber{\textit{ #2}}\thmnote{\textit{ (#3)}}}  

\theoremstyle{mystyle}
\newtheorem{theorem}{Theorem}
\newtheorem{remark}{Remark}
\newtheorem{lemma}{Lemma}
\newtheorem{corollary}{Corollary}
\newtheorem{definition}{Definition}

\makeatletter
\renewenvironment{proof}[1][\proofname]{\par
  \pushQED{\qed}
  \normalfont \topsep6\p@\@plus6\p@\relax
  \trivlist
  \item[\hskip\labelsep
        \itshape
    #1\@addpunct{:}]\ignorespaces
}{
  \popQED\endtrivlist\@endpefalse
}
\makeatother

\newcommand{\ra}[1]{\renewcommand{\arraystretch}{#1}}

\graphicspath{{Figures/}}
\allowdisplaybreaks
\begin{document}

\title{Uplink RSMA Performance Analysis with Rate Adaptation: A Stochastic Geometry Approach}
\author{Xinyi~Guo,~\IEEEmembership{Graduate Student Member, IEEE}, Li~You,~\IEEEmembership{Senior Member, IEEE},\\
Qiong~Liu,~\IEEEmembership{Member, IEEE},
Xiqi~Gao,~\IEEEmembership{Fellow, IEEE}, and Xiang-Gen~Xia,~\IEEEmembership{Fellow, IEEE}
\thanks{Xinyi Guo, Li You, and Xiqi Gao are with the National Mobile Communications Research Laboratory, Southeast University, Nanjing 210096, China, and also with the Purple Mountain Laboratories, Nanjing 211100, China (e-mail: xinyiguo@seu.edu.cn; lyou@seu.edu.cn; xqgao@seu.edu.cn).

Qiong Liu is with Sorbonne University, CNRS, LIP6, Paris 75006, France (e-mail: qiong.liu@sorbonne-universite.fr).

Xiang-Gen Xia is with the Department of Electrical and Computer Engineering, University of Delaware, Newark, DE 19716 USA (e-mail: xianggen@udel.edu).

Part of this work was presented at the 2025 IEEE WCNC~\cite{guo2025stochastic}.
}}

\maketitle

\begin{abstract}
Rate-splitting multiple access (RSMA) has emerged as a promising technique for efficient interference management in next-generation wireless networks. While most existing studies focus on downlink and single-cell designs, the modeling and analysis of uplink RSMA under large-scale deployments remain largely unexplored. On the basis of stochastic geometry (SG), this paper introduces a unified analytical framework that integrates finite modulation and coding scheme (MCS)-based rate adaptation. This framework jointly captures spatial interference coupling and discrete rate behavior to bridge theoretical tractability and practical realism. Within this framework, we derive tractable expressions for the conditional received rate (CRR), its spatial average, and higher-order statistics via the meta distribution, thereby quantifying both the mean and user-specific rate performance. Results show that the proposed unified framework not only generalizes existing non-orthogonal multiple access (NOMA) and orthogonal multiple access (OMA) analyses but also provides new insights into how discrete rate adaptation reshapes interference dynamics and fairness in dense RSMA-enabled networks.
\end{abstract}

\input{1-intro.tex}
\input{2-system}

\input{3-interference}
\input{4-linkLevel}

\input{5-symLevel}

\input{6-sim}
\input{7-conclusion}
\input{appendix}
\normalem
\bibliographystyle{IEEEtran}
\bibliography{reference}
\end{document}

%% file: 1-intro.tex
 \section{Introduction}
6G and beyond wireless networks are expected to support explosive growth in throughput, ultra-low latency, high reliability, and massive connectivity. These capabilities are essential for supporting emerging applications such as extremely reliable and low-latency communication (eURLLC), ultra-massive machine-type communication (umMTC), and the Internet of Everything (IoE)~\cite{wang2023road}. Meeting these stringent requirements calls for more efficient and flexible use of limited spectral resources, making the design of next-generation multiple access (NGMA) schemes a central research topic~\cite{ding2024next}. The main objective of NGMA is to support massive connectivity efficiently and flexibly within the constraints of available wireless radio resources. In this context, rate-splitting multiple access (RSMA) has emerged as a flexible multiple access (MA) framework that complements traditional schemes such as orthogonal multiple access (OMA) and non-orthogonal multiple access (NOMA) by providing a more general interference management strategy through message splitting~\cite{clerckx2023primer}. 

The core idea of RSMA is to split each user’s message into sub-messages that can be flexibly decoded by receivers through successive interference cancellation {(SIC)}. By allowing users to partially decode interference and partially treat it as noise based on its intensity, RSMA provides a flexible balance between interference cancellation and resource allocation. This flexibility makes RSMA a promising paradigm for future multi-user systems. 

While extensive research has investigated the benefits of RSMA in downlink {systems~\cite{clerckx2023primer}}, its analytical characterization in the uplink {large-scale networks remains largely unexplored, although the advantages of uplink RSMA in a single-cell system had been validated in~\cite{rimoldi1996rate} from the view of information theory}. In the downlink, interference can be managed through centralized precoding and coordinated power allocation at the transmitter. In contrast, the uplink poses distinct modeling challenges arising from two intrinsic factors: 1) independent user transmissions toward spatially distributed base stations (BS), leading to decentralized interference management{~\cite{tegos2025distributed,zamani2020optimizing}}, and 2) channel-dependent decoding order within each cell, where users are decoded according to their relative channel conditions rather than a fixed order as in the downlink~\cite{katwe2022rate,mao2022rate}. These difficulties become even more significant in dense network deployments, where a large number of small cells and users coexist in close proximity, intensifying inter-cell interference and spatial coupling. Therefore, a tractable analytical framework is required to characterize uplink RSMA performance under random and densely deployed conditions.

To capture the spatially coupled interference and randomness inherent in dense multi-cell networks, stochastic geometry (SG) provides a powerful analytical framework. By statistically modeling the spatial distribution of BSs and users, SG enables tractable evaluation of large-scale network performance beyond deterministic single-cell analyses.  
It enables analytical derivation of key performance metrics, such as coverage, mean rate, and meta distribution, thus offering insights into both the average and the variability of network behavior~\cite{haenggi2012stochastic,andrews2011tractable}.  
Recent works have successfully applied SG to study MA schemes, including OMA and NOMA, in dense cellular settings~\cite{liang2019non,salehi2018meta,tabassum2017modeling}.  
However, the analytical characterization of RSMA under random spatial configurations, particularly in the uplink, remains largely unexplored.

Building upon these motivations, this paper develops a tractable analytical framework for uplink RSMA networks in dense multi-cell environments using SG.  
The proposed model captures the spatial randomness of users and BSs, as well as interference coupling across cells.  
Furthermore, instead of assuming continuously adaptive transmission rates based on Shannon’s capacity formula, we incorporate a discrete modulation and coding scheme (MCS)-based rate adaptation, where transmission rates are determined by a finite set of Signal-to-Interference-plus-Noise Ratio (SINR) thresholds.   
This modeling provides a closer representation of practical transmission behavior and enables a direct comparison between discrete- and continuous-rate models in uplink RSMA networks. Overall, the proposed framework establishes a unified analytical foundation for evaluating uplink RSMA performance under spatial randomness and discrete rate adaptation.

\subsection{Related Work}
RSMA has emerged as one of the most widely studied NGMA techniques, owing to its ability to flexibly manage multi-user interference and enhance spectral efficiency across diverse network loads. Most studies have focused on the downlink implementation of RSMA in single-cell systems, where the transmitter can centrally coordinate power allocation~\cite{yang2021optimization,vu2025outage}, precoding~\cite{mishra2021rate,krishnamoorthy2022downlink}, and stream assignment~\cite{yang2024joint,lyu2024rate}. Besides, authors in \cite{mao2022rate} have shown how RSMA can generalize and encompass space division multiple access (SDMA), OMA, NOMA, and multi-casting as special cases, demonstrating its potential advantages over traditional methods in certain network conditions. 

However, despite these advances, research on uplink RSMA and multi-cell RSMA-enhanced networks remains relatively limited. Early studies~\cite{rimoldi1996rate} showed that uplink RSMA can achieve points on the capacity region without time-sharing, thereby simplifying implementation compared to NOMA. Subsequent works further studied its performance in terms of sum-rate maximization~\cite{yang2020sum}, user fairness enhancement~\cite{liu2020rate}, and energy efficiency maximization~\cite{jiang2023rate} under various system settings. In the context of multi-cell networks, research has explored the potential of RSMA to mitigate inter-cell interference and enhance resource allocation, though most studies remain focused on downlink systems. For example, the coordinated RSMA scheme proposed in~\cite{ha2020coordinated} mitigates inter-cell interference and inter-user interference in a two-cell downlink system. In~\cite{mao2019rate}, the authors investigated RSMA in downlink coordinated multi-point joint transmission networks via beamformer design.  

Despite these advances, existing RSMA research has primarily focused on deterministic or single-cell configurations. 
Stochastic geometry offers a tractable analytical framework to capture spatial randomness in large-scale multi-cell networks, where interference and user distributions vary across space~\cite{haenggi2012stochastic,haenggi2015meta}. 
SG also enables the study of statistical dependencies, such as interference correlation and spatial coupling, that are difficult to capture in deterministic analyses. Recent studies have leveraged SG to analyze the performance of MA schemes in dense networks.
For instance, it has been used to derive transmission success probability of uplink NOMA systems~\cite{liang2019non}, evaluate outage probability and average rate in multi-cell systems~\cite{kusaladharma2019outage}, and study the area spectral efficiency of multiple-input multiple-output (MIMO)-NOMA networks~\cite{chen2020performance}. Despite progress in SG-based analyses of NOMA, the study of uplink RSMA within the SG framework remains at an early stage. To the best of our knowledge, only a few works, e.g.,~\cite{zhu2023rate}, have studied RSMA under spatial randomness, focusing on downlink  networks and demonstrating its performance gains over NOMA.

Furthermore, most SG-based analyses of MA schemes assume continuously adaptive transmission rates, following Shannon’s capacity law, for analytical convenience. In contrast, practical wireless systems rely on discrete MCS, where achievable rates are determined by a finite set of SINR thresholds. Our previous work~\cite{guo2025stochastic} analyzed the impact of MCS adaptation in OMA systems. Several studies have investigated the MCS selection for RSMA based communication~\cite{dizdar2020rate,mishra2021rate}. However, these studies are confined to link-level or single-cell settings and overlook the spatial coupling effects present in large-scale multi-cell networks. In this paper, we bridge these gaps by incorporating MCS-based discrete rate adaptation into a SG framework for uplink RSMA, thereby providing a tractable yet realistic analytical model for dense cellular systems.

\subsection{Motivation and Contributions}
Existing analytical studies on RSMA have predominantly focused on single-cell or downlink configurations, often assuming idealized continuous rate models based on Shannon capacity. However, uplink RSMA-enhanced multi-cell networks in real-world environments introduce distinct challenges that remain analytically unresolved. First, interference arises in a decentralized manner from spatially distributed users, making tractable modeling of inter-cell coupling non-trivial. Second, practical systems employ discrete MCS, where achievable rates depend on finite SINR thresholds rather than continuous adaptation. These characteristics motivate the need for a tractable framework that captures both spatial randomness and discrete rate adaptation in large-scale uplink RSMA networks.

To bridge this gap, this paper develops an SG-based analytical framework for uplink RSMA systems with finite MCS adaptation. The proposed approach integrates the spatial distribution of BSs and users with the rate-splitting  mechanism, enabling tractable system-level analysis of uplink RSMA networks. The main contributions are summarized as follows:
\begin{itemize}
   \item \textit{SG-based modeling of uplink RSMA with finite MCSs.} We establish an SG framework that integrates the spatial distribution of users and BSs with the uplink rate-splitting transmission.  
   The model incorporates discrete MCS-based rate adaptation and introduces the conditional received rate (CRR) as a key metric for tractable performance evaluation.
    \item \textit{Tractable analytical characterization of spatially averaged rates.} We derive compact expressions for the spatially-averaged CRR and its Shannon-based upper bound (average achievable rate), quantifying the practical–ideal rate gap in uplink RSMA. 
    \item \textit{Moment- and meta distribution- based analysis.} We extend the framework beyond spatial averages by deriving the moments and meta distribution of the CRR.
    This analysis reveals the variability and distribution of user rates across the network, providing a more comprehensive statistical understanding of uplink RSMA.
    \item \textit{Unified analytical framework for MA schemes.} The proposed model provides a unified analytical basis that accommodates various MA techniques, including NOMA and OMA, through appropriate parameter selection.
    This generality allows systematic performance benchmarking and facilitates a fair comparison of transmission rates and their variability across different access paradigms. 
\end{itemize}

The rest of this paper is organized as follows.
Section~\ref{sec:sym} introduces the system model and key performance metrics. 
Section~\ref{sec_link} analyzes the link distribution and interference. 
Section~\ref{sec:spatial} derives the CRR and spatially-averaged performance,
while Section~\ref{sec:fine} focuses on high-order statistics, including the moments and meta distribution of CRR.
Section~\ref{sec:sim} provides numerical validations and comparative evaluations with multiple MA schemes, and Section~\ref{sec:conclusion} concludes the paper.

%% file: 2-system.tex
\section{System Model}
\label{sec:sym}
\subsection{Network Topology}
We consider a single-tier uplink cellular RSMA-assisted network, where BSs are spatially distributed according to a homogeneous Poisson point process (PPP) $\Phi_{\text{BS}}$ with intensity $\lambda_{\text{BS}}$ in $\mathbb{R}^2$. User equipments (UE) are independently distributed as a homogeneous PPP $\Phi_{\text{UE}}$ with intensity $\lambda_{\text{UE}}$. {Each BS selects $N$ UEs for service during each resource block under the assumption of $\lambda_{\text{UE}} \gg \lambda_{\text{BS}}$, such that each BS has more than $N$ UEs within its coverage area.} UEs transmit signals to their serving BSs using RSMA. 

Each selected UE associates with its nearest BS, thereby inducing a standard Voronoi tessellation of the plane. Within each Voronoi cell, $N$ UEs are uniformly and independently selected to transmit simultaneously. The resulting scheduled user process is denoted by $\Phi_{\rm u} \triangleq \{U(V(y)) \mid y \in \Phi_{\text{BS}}\}$, where $V(y)$ is the Voronoi cell of BS $y$ and $U(V(y))$ represents the set of selected UEs within $V(y)$~\cite{haenggi2017user}. 

\begin{figure}
 \centering
 \includegraphics[width=0.38\textwidth]{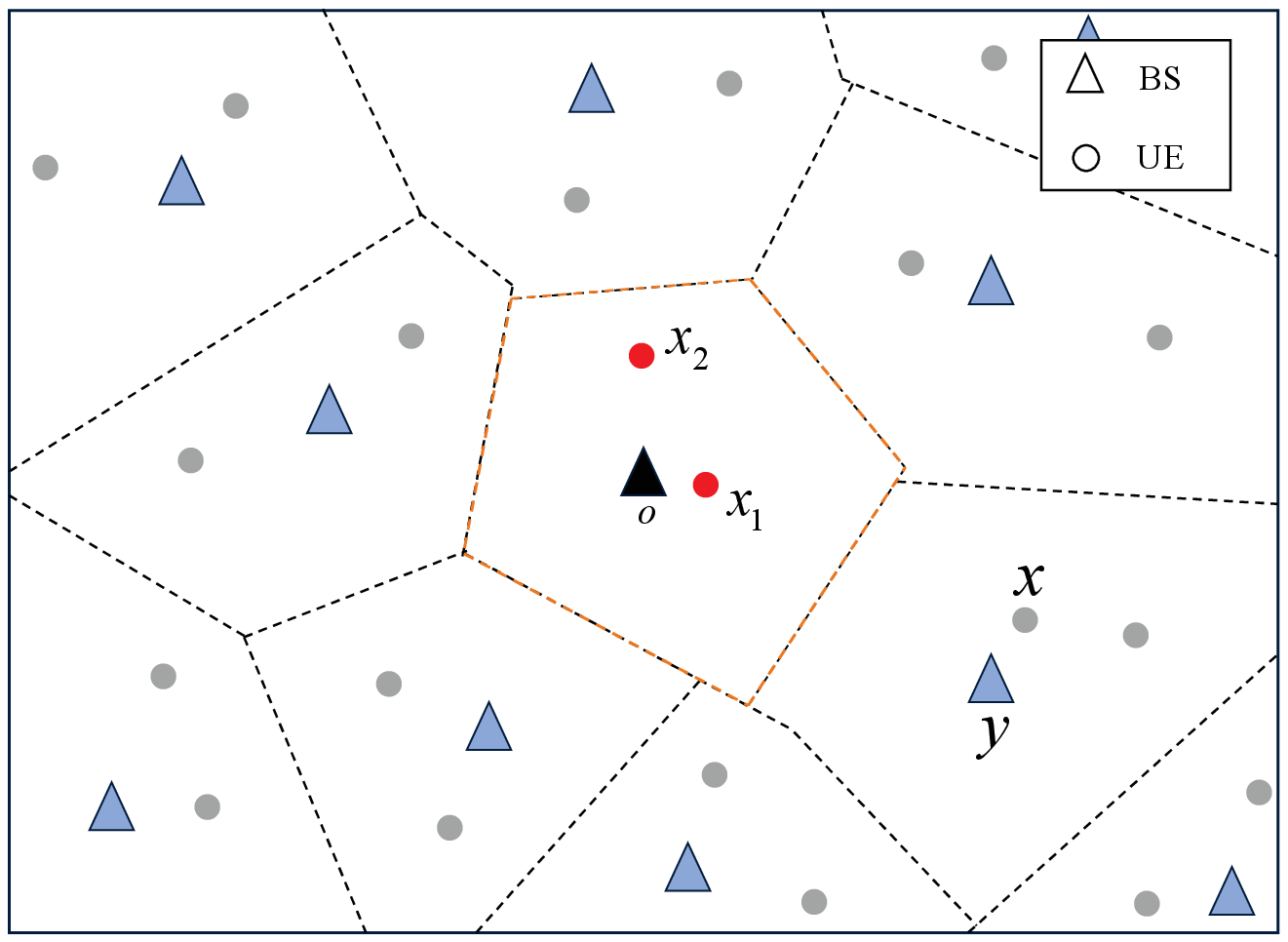}
 \caption{A realization of network topology with $N=2$, where $x_1$ and $x_2$ denotes the 1st-ranked and 2nd-ranked typical UE, respectively.}
 \label{fig_realization}
\end{figure}

By Slivnyak’s theorem~\cite{haenggi2012stochastic}, we focus on a typical BS, which can be assumed to be located at the origin $o \in \mathbb{R}^2$ without loss of generality. Accordingly, the corresponding cell and its associated UEs are referred to as the typical cell and the typical UEs, respectively. We denote $\Phi_{\rm I}$ as the point process of UEs from other cells, i.e., inter-cell interferers. In other words, for each of its realization and for a region $A\subset \mathbb{R}^2$, $\Phi_{\rm I}(A)$ is a set of indices of the UEs from other cells in region $A$. We assume the association distances between the typical BS and its $N$ typical UEs, denoted by $R_{(n)}$ for $ n=1,\cdots, N$, are independent and identically distributed (i.i.d.) with probability density function (PDF) $f_{R}(r)$ and cumulative distribution function (CDF) $F_{R}(r)$. Let $R_1\leq\cdots\leq R_N$ be the ordered random variables of $R_{(n)}$, $ n=1,\cdots,N$. Similarly, $D_x$ denotes the distances between the typical BS and each inter-cell interferer $x\in\Phi_{\rm I}$.  Fig. \ref{fig_realization} illustrates a realization of network topology with $N=2$. 

\subsection{Signal-to-Interference-plus-Noise Ratio}
We consider a general uplink RSMA scheme~\cite{clerckx2023primer}. In the cell served by BS $y\in\Phi_{\text{BS}}$, message $W_n^y$ for its $n$-th ranked associated UE is split into two sub-messages $w_{n,1}^y$ and $w_{n,2}^y$. They are encoded into independent signals $s_{n,1}^y$ and $s_{n,2}^y$ with unit power. Thus, the transmit signal is given by
\begin{align}
  v_n^{y} =\sqrt{P_{1}}s_{n,1}^{y}+\sqrt{P_{2}}s_{n,2}^{y},
\end{align}
where $P_{1}$ and $P_{2}$ are transmit powers of $s_{n,1}^{y}$ and $s_{n,2}^{y}$, respectively. {We assume that each UE has the same total transmit power $P_{1}+P_{2}=P$, and denote the power allocation fraction as $\beta \in [0,1]$, which is the same for every UE.} Thus, we have $P_{1} = \beta P$ and $P_{2} = (1-\beta) P$.
We then rewrite the transmit signal as
\begin{align}
 v_{n}^{y} =\sqrt{P}( \sqrt{\beta} s_{n,1}^{y}+\sqrt{1-\beta} s_{n,2}^{y}).
\end{align}
Within each resource block, all UEs' signals are transmitted via a multiple access channel (MAC), and the received signal at the typical BS $o$ is given by
\begin{align}
 Y\!\!\!=\!\!\sqrt{P}\sum_{n=1}^{N}\!{h_{n}^{o}}(\!\sqrt{\beta}s_{n,1}^{o}\!\!+\!\!\sqrt{\!1\!-\!\beta}s_{n,2}^{o})\!+\!\!\!\!\!\sum_{y\in\Phi_{\text{BS}}\setminus o}\sum_{n=1}^{N}{h_{n}^{y}}v_{n}^{y}\!+\!z_{o},
\end{align}
where $h_{n}^{y}$ is the channel vector between BS $y$ and its associated $n$-th ranked UE, $z_{o}$ is the additive noise at BS $o$. In this paper, we consider perfect distance-based SIC~\cite{tabassum2017modeling}. The BS determines the decoding order by sorting the associated UEs by distance, and we assume error-free decoding. Additionally, we assume that each UE's sub-messages are decoded successively due to the dominant role of distance in SIC and the co-location of sub-messages.
Consider an interference and noise-limited network, we can obtain the received SINR of the typical BS $o$ from its $n$-th ranked typical UE's two sub-signals, as follows 
\begin{align}
&\text{SINR}_{n,1}^{\text{RS}}=\frac{|h_{n}^{o}|^{2}\beta}{b_n|h_{n}^{o}|^{2}(1-\beta)+I_{n}^{\text{intra}}+I^{\text{inter}}+\tilde{\sigma}^2},\notag\\
&\text{SINR}_{n,2}^{\text{RS}}=\frac{|h_{n}^{o}|^{2}(1-\beta)}{(1-b_n)|h_{n}^{o}|^{2}\beta+I_{n}^{\text{intra}}+I^{\text{inter}}+\tilde{\sigma}^2},
\end{align}
where ``RS" denotes RSMA. $b_n \sim \text{Bernoulli}\ (q)$ models the randomness in the decoding order of the UE’s sub-messages. Here, $q$ is referred to as the inter-UE decoding order factor. $\tilde{\sigma}^2$ is the normalized additive noise power with $\tilde{\sigma}^2=\frac{\sigma^2}{P}$. $I_{n}^{\text{intra}}=\sum_{i=n+1}^{N}|h_{i}^{o}|^{2}$ and $I^{\text{inter}}=\sum_{y\in \Phi_{\text{BS}}\setminus{o}}\sum_{n=1}^{N} |h_{n}^{y}|^{2}$ are normalized intra-cell and inter-cell interference of the $n$-th ranked typical UE. Actually, $b_n|h_{n}^{o}|^{2}(1-\beta)$ and $(1-b_n)|h_{n}^{o}|^{2}\beta$ are normalized inter-UE interference for sub-message $1$ and sub-message $2$, respectively. 

For ease of analysis, we redefine $|h_{n}^{o}|^{2}=H_{n}^{o}R_{n}^{-\eta}$, $|h_{i}^{o}|^{2} =H_{i}^{o}R_{i}^{-\eta} $ and $|h_{n}^{y}|^{2} = H_{x}D_{x}^{-\eta}$ to distinguish the large-scaling and small-scaling fading, where $\eta$ denotes the path loss exponent.{~It is noteworthy that only links indexed by $i>n$ contribute to intra-cell interference under the distance-based decoding strategy. However, the range of $i$ is actually $\{i|i\in\mathbb{N},1\leq i\leq N,i\neq n\}$. When it comes to interference, we assume by default that $i>n$.}{ Recall that $R_{n}$, $R_{i}$ and $D_{x}$ are the link distances between BS $o$ and its $n$-th ranked associated UE, $i$-th ranked associated UE and inter-cell interfer $x\in \Phi_{\rm I}$, respectively}. In addition, we assume $H_{n}^{o}$, $H_{i}^{o}$, and $H_x$ are corresponding channel gains of these links and follow an i.i.d. exponential distribution with unit parameter, i.e., $\exp(1)$. Hence, we rewrite the SINR expressions as
\begin{align}
\label{eq_SINR_RS}
& \text{SINR}_{n,1}^{\text{RS}} = \frac{\beta H_{n}^{o} R_{n}^{-\eta}}{b_n H_{n}^{o} R_{n}^{-\eta}(1-\beta)+I_{n}^{\text{intra}}+I^{\text{inter}}+\tilde{\sigma}^2},\notag\\
& \text{SINR}_{n,2}^{\text{RS}} = \frac{(1-\beta) H_{n}^{o} R_{n}^{-\eta}}{(1-b_n) H_{n}^{o} R_{n}^{-\eta}\beta+I_{n}^{\text{intra}}+I^{\text{inter}}+\tilde{\sigma}^2},
\end{align}
where $I^{\text{inter}}\!=\!\sum_{x\in \Phi_{\rm I}}H_x D_x^{-\eta}$, $I_{n}^{\text{intra}}\!=\!\sum_{i=n+1}^{N}\!H_{i}^{o}R_{i}^{-\eta}$.
\begin{remark}
 NOMA is a special case of RSMA when the power allocation fraction $\beta=0$ or $1$, thus the corresponding SINR expression for the $n$-th ranked typical UE is reduced to
\begin{align}
 \label{eq_SINR_NO}
 {\text{SINR}_{n}^{\text{NO}}=\frac{H_{n}^{o} R_{n}^{-\eta}}{\sum_{i=n+1}^{N}H_{i}^{o} R_{i}^{-\eta}+\sum_{x\in \Phi_{\rm I}} H_{x} D_x^{-\eta}+\tilde{\sigma}^2}},
\end{align}
Here, ``NO'' denotes the NOMA scenario. In the OMA scenario, only $N=1$ UE is served per resource block without intra-cell interference. The SINR of the sole UE is given by
\begin{align}
\label{eq_SINR_OMA}
 \text{SINR}^{\text{OMA}}=\frac{H_o R^{-\eta}}{\sum_{x\in \Phi_{I}}H_x D_x^{-\eta}+\tilde{\sigma}^2},
\end{align}
where $R$ denotes the typical link distance and $H_o$ represents the channel gain of this link.
\end{remark}
Note that the SINR depends on $R_n$, $R_i$, and $D_x$, which capture the spatial randomness of the network. In Section \ref{sec_link}, we further model and analyze these link distances and the resulting interference (inter-UE, intra-cell, and inter-cell).
\newcolumntype{L}{>{\hspace*{-\tabcolsep}}l}
\newcolumntype{R}{c<{\hspace*{-\tabcolsep}}}
\definecolor{lightblue}{rgb}{0.93,0.95,1.0}
\begin{table}[!t]
\captionsetup{font=footnotesize}
\caption{List of Notations}
\label{tb:simulation parameters}
\centering
\ra{1.6}
\scriptsize
	\begin{tabular}{LR}
		\toprule
		Symbol &  Definition\\
		\midrule
		\rowcolor{lightblue}
		$\lambda_{\text{BS}}/\lambda$ & BS intensity  \\
        $\Phi_{\text{BS}}, \Phi_{\text{UE}}, \Phi_{\rm{I}}$ & BS, UE, and Inter-cell interfering UE processes \\
        \rowcolor{lightblue}
        $N$ & Number of UEs served by each BS in a resource block \\
        $R_{n}$ & Distance from the  $n$-th ranked typical UE to the typical BS \\
        \rowcolor{lightblue}
        $D_x$ & Distance from the inter-cell interferer $x$ to the typical BS \\
        $H_{n}^{o}$ & Small-scale fading gain of the $n$-th typical link \\
        \rowcolor{lightblue}
        $H_x$ & Small-scale fading gain between $x$ and the typical BS\\
        $\eta$ & Path loss exponent\\
        \rowcolor{lightblue}
        $\tilde{\sigma}^2$ & Normalized noise power \\
        $\beta$ & Power allocation fraction in RSMA\\
        \rowcolor{lightblue}
        $q$ & Inter-UE decoding oder factor\\
        \rowcolor{lightblue}
        $\Theta=\{\theta_{i_{\mathrm{CQI}}}\}_{1}^{M}$ & A set of SINR thresholds\\
        $\{r_{i_{\mathrm{CQI}}}\}_{0}^{M}$ & Rates for disjoint SINR regions\\
        \bottomrule
	\end{tabular}
\end{table}

\subsection{MCS Adaptation and Performance Metrics}
In modern wireless communication networks, MCS adaptation serves as a practical implementation to approach the Shannon theoretical limit. Specifically, the Shannon limit defines the maximum transmission rate for reliable communication over a channel susceptible to random bit errors~\cite{6773024}. Given the channel capacity $C$, the achievable rate $\bar{R}$ for any practical communication system satisfies
\begin{align}
\label{eq:achi_R}
 \bar{R}=BW\cdot \log_2 (1+\mathrm{SINR})\leq C,
\end{align}
where $BW$ is the system bandwidth. 

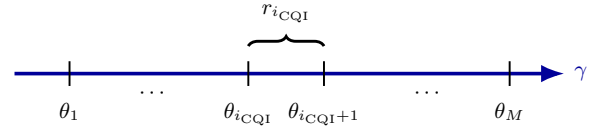
\begin{figure}[t]
\centering
\resizebox{0.9\linewidth}{!}{
\begin{tikzpicture}[>=Latex, line width=1.2pt, font=\small]
  \draw[->, line width=1.5pt, color=blue!60!black] (0,0) -- (8,0) node[right]{$\gamma$};

\foreach \x/\label in {
  0.8/{\(\theta_1\)},
  3.4/{\(\theta_{i_{\mathrm{CQI}}}\)},
  4.5/{\(\theta_{i_{\mathrm{CQI}}+1}\)},
  7.2/{\(\theta_M\)}
}{
  \draw[thick] (\x,0.18) -- (\x,-0.18);
  \node[below=3pt] at (\x,-0.18) {\label};
}
\node at (2,-0.28) {\(\cdots\)};
\node at (6.0,-0.28) {\(\cdots\)};
  \draw[decorate,decoration={brace, amplitude=4pt}] (3.4,0.4) -- (4.5,0.4)
    node[midway,above=6pt]{\(\,r_{i_{\mathrm{CQI}}}\,\)};
\end{tikzpicture}
}
\caption{A multi-level MCS strategy.}
\label{fig_MCS_r}
\end{figure}

According to the MCS adaptation mechanism, each MCS is associated with a specific SINR threshold. Adaptive modulation and coding is implemented via sounding reference signals, enabling the BS to estimate the SINR and select a suitable MCS level, typically indexed by a channel quality indicator (CQI)~\cite{lu2021stochastic}. To formalize this mechanism, we model a multi-level MCS strategy by dividing the SINR range into $M+1$ non-overlapping intervals defined by the threshold set $\Theta = \{\theta_1, \dots, \theta_M\}$ with $\theta_1 < \cdots < \theta_M$. For completeness, we set $\theta_0 = -\infty$ and $\theta_{M+1} = +\infty$. Each interval $[\theta_m, \theta_{m+1})$ is mapped to an MCS level $i_{\text{CQI}} = m \in \{0,1,\dots,M\}$, with $r_0 = 0 < r_1 < \cdots < r_M$ denoting the associated rate levels. The dividing strategy is illustrated in Fig. \ref{fig_MCS_r}.
Therefore, for a given SINR value $\gamma$, the transmission rate is given by
\begin{align}
\label{eq_r_mcs}
 R_{\text{MCS}}(\gamma) =\sum_{i_{\text{CQI}}=0}^{M} r_{i_{\text{CQI}}} \cdot \mathds{1}\left(\theta_{i_{\text{CQI}}} \leq \gamma < \theta_{i_{\text{CQI}+1}}\right),
\end{align}
where $\mathds{1}(\cdot)$ denotes the indicator function. We further define conditional received rate and spectral efficiency as follows.
\begin{definition}[{Conditional received rate (CRR) and spectral efficiency (SE)}]
\label{def_CRR}
The CRR of the typical BS from the $n$-th ranked typical UE is given by 
\begin{align}
\bar{r}_{n}(\Phi_{\text{BS}},\Phi_{\rm u})=\mathbb{E}[R_{\text{MCS}}(\gamma_{n})|\Phi_{\text{BS}},\Phi_{\rm u}].
\end{align}
In additon, the conditional SE for the $n$-th ranked UE with received SINR $\gamma_n$ is defined as follows~\cite{martin2017interference}
\begin{align}
\text{SE}_{\text{MCS}}(\gamma_n)=\sum_{i_{\text{CQI}}=0}^{M}\text{SE}_{i_{\text{CQI}}}\mathds{1}\left(\gamma_n \in\left[\theta_{i_{\text{CQI}}},\theta_{i_{\text{CQI}}+1}\right)\right),
\end{align}
where \( \text{SE}_{i_{\text{CQI}}} \)  is the SE (bit/s/Hz) corresponding to the $i_{\text{CQI}}$-th MCS level, and $\text{SE}_{\text{MCS}}=r_{i_{\text{CQI}}}/BW$.
\end{definition}

%% file: 3-interference.tex
\section{Link Distance and Interference Analysis}
\label{sec_link}
\begin{figure}
    \centering
    \includegraphics[width=0.38\textwidth]{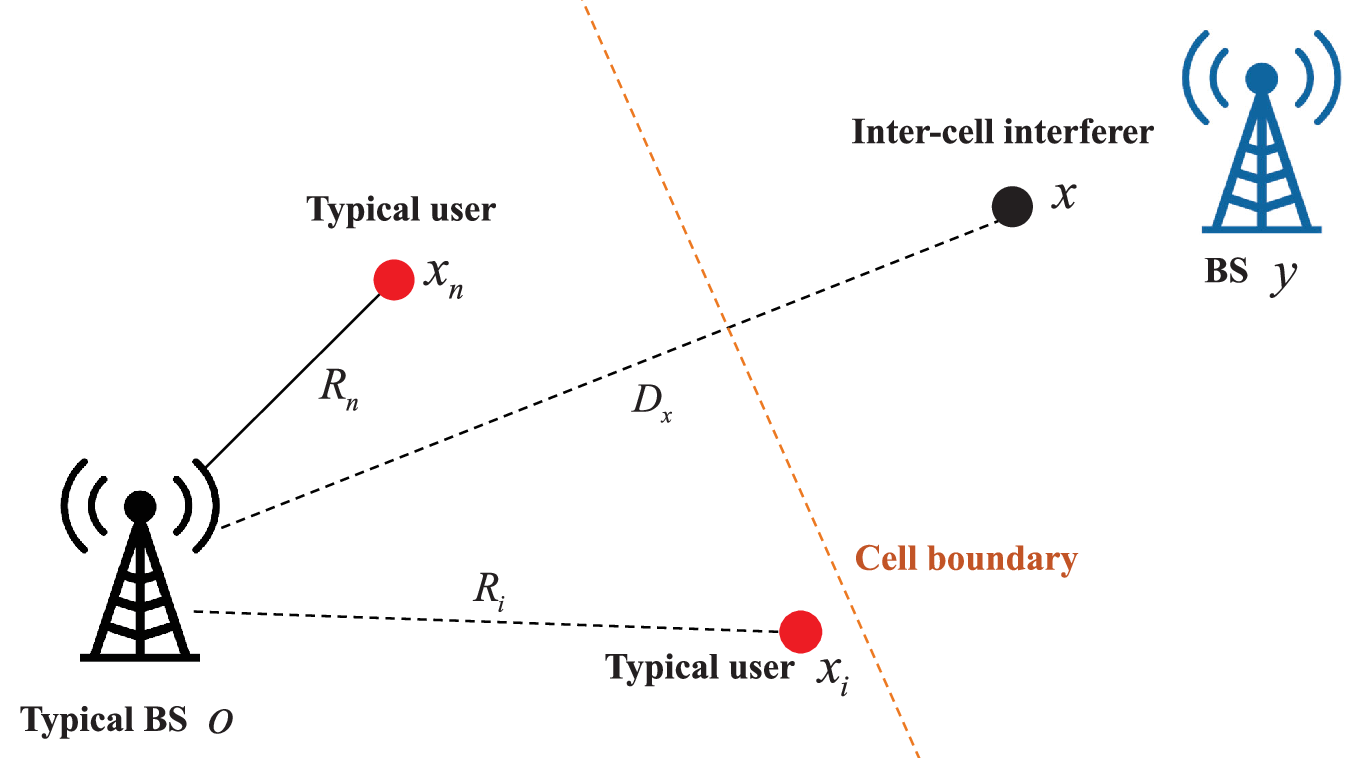}
\caption{Three distances in uplink interference: 
$R_n$ (scheduled UE), 
$R_i$ (intra-cell interferers), 
and $D_x$ (inter-cell interferers).}
    \label{fig_distance_diagram}
\end{figure}

This section analyzes uplink interference by characterizing intra- and inter-cell UE distributions, along with their distance-dependent path loss and interference effects. Considering the typical BS located at the origin, we focus on three key distances,  $R_{n}$, $R_{i}$, and $D_{x}$, as illustrated in Fig. \ref{fig_distance_diagram}.

\subsection{Distribution of Link Distance $R_n$}
We first analyze the distance between the typical BS and its $n$-th ranked associated UE, which directly determines the desired signal power. {When UEs follow a homogeneous PPP, the contact distance distribution is exactly a Rayleigh distribution with mean $\frac{1}{2\sqrt{\lambda_{\text{BS}}}}$, derived from the void probability of a circular region with radius smaller than the contact distance. Notably, this distance distribution depends on cell size, which is determined by $\lambda_{\text{BS}}$. However, the spatial distribution of UEs is actually inhomogeneous; thus, we approximate it by a Rayleigh distribution with mean $1/\sqrt{5\lambda_{\text{BS}}}$~\cite{wang2017meta}, accounting for the correlation factor. To simplify notation, we use $\lambda$ instead of $\lambda_{\text{BS}}$ to denote BS intensity. The PDF and CDF of the unordered link distances $R_{(k)}$ for $k=1,\cdots,N$ are given as follows }
\begin{align}
\label{eq:fr}
 f(r)\!=\!2B_1 \pi \lambda r e^{\!-B_1 \lambda \pi r^2}, F(r)\!=\!1\!-\!e^{\!-B_1\lambda \pi r^2},
\end{align}
where $B_1=\frac{5}{4}$. Therefore, the distribution of the desired distance $R_{n}$ on the basis of order statistics~\cite{david2004order} is 
\begin{align}
 f_{R_{n}}(r)=\frac{2B_1 \lambda \pi r\left(1-e^{-B_1 \lambda \pi r^{2}}\right)^{n-1}\left(e^{-B_1 \lambda \pi r^{2}}\right)^{N-n+1}}{ B(N-n+1, n)}, 
\label{eq_f_Rm}
\end{align}
where $B(\cdot,\cdot)$ is the Beta function. 

\subsection{Distribution of $R_i$ and Intra-Cell Interference}
In RSMA systems, multiple users are simultaneously scheduled by the same BS within a resource block, leading to intra-cell interference among co-scheduled UEs. The distances $R_i$, $i\neq n$, determine the large-scale fading of these interfering links. Thus, they play a central role in characterizing the intra-cell interference field of the $n$-th ranked typical UE. {The ordered distance random variables $R_n$ and $R_i$ for $i\neq n$ are mutually dependent. Hence, modeling the dependency between $R_i$ and $R_n$ is critical.} 

Given that we adopt a distance-based decoding strategy, our primary focus lies in the regime where $i$ exceeds $ n$. Hence, we derive the joint PDF of $R_{n+1},\cdots, R_N$ conditioned on $R_n=a$ as follows~\cite{david2004order}
\begin{align}
\label{eq:cf_r}
     &f_{R_{n+1}, \cdots, R_{N}}(r_{n+1}, \cdots, r_{N}|{R_n=a})\notag\\
     =&\frac{(N\!-n)!\prod_{i=n+1}^{N}f(r_i)}{[1\!-\!F(a)]^{N\!-\!n}},\ a\leq r_{n+1}\leq \cdots \leq r_N,
\end{align}
where $f(\cdot)$ and $F(\cdot)$ follow~\eqref{eq:fr}.

\subsection{Distribution of $D_x$ and Inter-Cell Interference}
\label{sec:inter}
In multi-cell dense networks, the typical BS is subject to uplink inter-cell interference from UEs in other cells, denoted as $\Phi_{\rm{I}}$. Due to the correlation between BSs and their associated UEs, $\Phi_{\rm{I}}$ follows a non-stationary distribution {whose exact statistics are intractable}~\cite{haenggi2017user}. {Thus, we resort to suitable approximations to strike a good balance between the tractability of subsequent derivations and accuracy.} In particular, the interference process $\Phi_{\rm{I}}$ when $N=1$ was analyzed in our earlier work~\cite{guo2025stochastic}, 
where it was approximated by an inhomogeneous PPP. However, the interference field is more involved when $N>1$. 
Simultaneous multi-user scheduling introduces clustering and correlation among interferers. To capture these effects, we analyze $\Phi_{\rm{I}}$ through its pair correlation function. For notational simplicity, the inter-cell distance $D_x$ is denoted by $r$ throughout this subsection.

\subsubsection{Pair Correlation Function and K Function}
\begin{figure}
    \centering
    \includegraphics[width=0.38\textwidth]{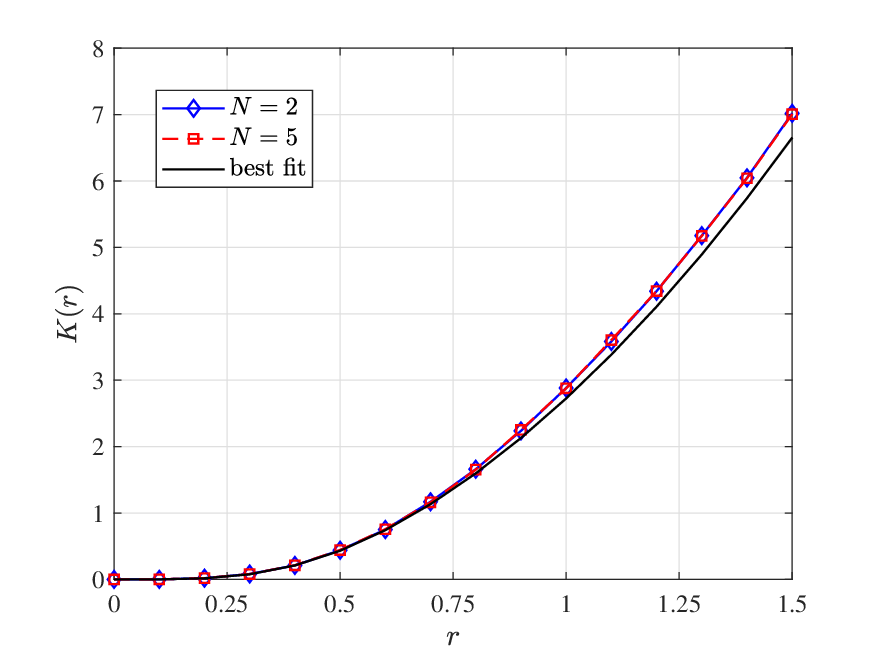}
    \caption{{The simulated $K$ functions of $\Phi_{\rm I}$ corresponding to $N=2$ and $N=5$ selected UEs, along with the best-fit theoretical $K$ function shown in \eqref{eq:g1}.}}
    \label{fig_K}
\end{figure}

We first use the pair correlation function to characterize the spatial dependence between the typical BS and the interfering UEs. For motion-invariant BS and UE point processes, the BS-UE pair correlation function $g_{\lambda}(r)$ is defined as~\cite{haenggi2012stochastic} 
\begin{align}
\label{equ:gr}
    g_{\lambda}(r)\!\triangleq\!\frac{1}{2\pi r}\!\frac{\mathrm{d}}{\mathrm{d}r}K(r)\!=\!\frac{1}{2\pi r}\!\frac{\mathrm{d}}{\mathrm{d}r}\!\left(\!\frac{1}{N\lambda}\mathbb{E}_{o}^{!}[\bar{N}_{\Phi_\mathrm{I}}\left(b\left(o,r\right)\right)]\!\right)\!,
\end{align}
where $r$ is the BS-UE pair distance, $K(r)$ is Ripley's $K$ function, and $b(o,r)$ represents the ball with radius $r$ centered at the origin $o$. $\bar{N}_{\Phi_{\rm I}}$ is a random counting measure associated with $\Phi_{\rm I}$, satisfying $\bar{N}_{\Phi_{\rm I}}(A)=|\Phi_{\rm I}(A)|\in \mathbb{N}\cup\{0\},A\subset \mathbb{R}^2$, where $|\cdot|$ denotes the cardinality of a set. Thus, $\bar{N}_{\Phi_\mathrm{I}}\left(b\left(o,r\right)\right)$ denotes the number of interfering UEs residing in  $b\left(o,r\right)$, which is a random random variable. $\mathbb{E}_{o}^{!}$ denotes the reduced Palm expectation, where the subscript $o$ indicates that the typical BS is known to exist at the origin $o$, and the superscript $!$ denotes the removal of the UEs associated with the typical BS. $\mathbb{E}_{o}^{!}[\bar{N}_{\Phi_\mathrm{I}}\left(b\left(o,r\right)\right)]$ denotes the intensity measure of $\Phi_{\rm{I}}$, i.e., the expected number of interfering UEs in the region $\left(b\left(o,r\right)\right)$, given the typical BS at $o$. When $N=1$ (i.e., in OMA systems), the interference field $\Phi_{\rm{I}}$ can be approximated by an inhomogeneous PPP with intensity $\lambda_{\rm{I}}(r)=\lambda\left(1\!-\! e^{-\frac{12}{5}\lambda \pi r^2}\right)$, and the pair correlation function {and $K$ function are} reduced from \eqref{equ:gr} to
\begin{align} 
g_{\lambda}(r)\!=\!1\!-\!e^{-(12/5)\lambda\pi r^2},\  K(r)\!=\!\pi r^2\!+\!\frac{5}{12}e^{-\frac{12}{5}\pi r^2}\!-\!\frac{5}{12}.
\label{eq:g1}
\end{align} 
To validate this model {for $N>1$}, we simulate the $K$ functions for $N=2$ and $N=5$, and compare results with that of \eqref{eq:g1}, as shown in Fig.~\ref{fig_K}. We observe that the $K$ function  remains nearly invariant with respect to  $N$, since the average number of interferers seen by the typical BS scales linearly with $N$. {Moreover, }it can be concluded that the BS-UE pair correlation function exhibits the invariance property {with respect to the UE number $N$ in each cell according to \eqref{equ:gr}}. Based on this observation, we investigate the approximation of $\Phi_{\rm{I}}$ by employing two point process models in the following analysis. 

\subsubsection{Inter-Cell UE Point Process Approximation} 
Due to the complex spatial randomness exhibited by the point process $\Phi_{\rm I}$, and considering the above observations, we investigate two tractable point processes, which are a Poisson cluster process (PCP) $\Phi_{\rm A}$ and an inhomogeneous PPP $\Phi_{\rm B}$~\cite{salehi2018meta}. The subscripts A and B are used to distinguish these two models.
 
\textbf{Model A:} {Building on the observation that the average number of interfering UEs increases linearly with $N$ (selected UE numbers in each cell), we} model $\Phi_{\rm I}$ as a PCP $\Phi_{\rm A}$ generated from a parent point process, which is an inhomogeneous PPP. Specifically, we assume that in each interfering cell, there are $N$ UEs uniformly distributed in the disk centered at a parent point $x_{\rm p}$ residing in this cell, {with a given radius. To avoid prohibitively complex analysis, we set the radius to zero for each cluster, i.e., a co-location case here.} These parent points are assumed to form an inhomogeneous PPP seen from the typical BS, denoted as $\Phi_{\rm A}^{\rm p}$, with intensity {$\lambda_{\rm A}^{\rm p}(r)=\!\!\lambda\cdot g_{\lambda}(r)$ and $g_{\lambda}(r)$} follows \eqref{eq:g1}. Thus, the interfering UEs seen from the typical BS are modeled as $\Phi_{\rm A}\!\!=\!\cup_{x_{\rm p}\!\in\! \Phi_{\rm A}^{\rm p}} \mathcal{N}^{x_{\rm p}}$, where $\mathcal{N}^{x_{\rm p}}$ denotes the set of interfering UEs surrounding the parent point $x_{\rm p}$. While the cluster constraint of $\Phi_{\rm A}$ may not match the specific distribution characteristics of the practical interfering UE point process $\Phi_{\rm I}$ within each cell (e.g., interfering UEs are not randomly distributed only in a disk region), from the perspective of the typical BS, $\Phi_{\rm A}$ still provides an effective approximation to {the distribution} of $\Phi_{\rm I}$.    

\textbf{Model B:}
Also based on the observation, we may directly model $\Phi_{\rm I}$ as an inhomogeneous PPP $\Phi_{\rm B}$ with its intensity function defined as {$\lambda_{\rm B}(r)=N\lambda\cdot g_{\lambda}(r)$}, where $g_{\lambda}(r)$ follows \eqref{eq:g1} {and $N$ denotes the selected UE numbers in each cell for each resource block}. On one hand, the density of interfering UEs within the extremely short-distance range of the typical BS decreases sharply, characterizing the repulsion between the near interferers and the typical BS. On the other hand, as $r$ approaches infinity, the interfering density gradually converges to a constant $N\lambda$. This property aligns with the physical intuition that the interferer distribution tends to be uniform in far-field scenarios.

\begin{figure}
    \centering
\includegraphics[width=0.38\textwidth]{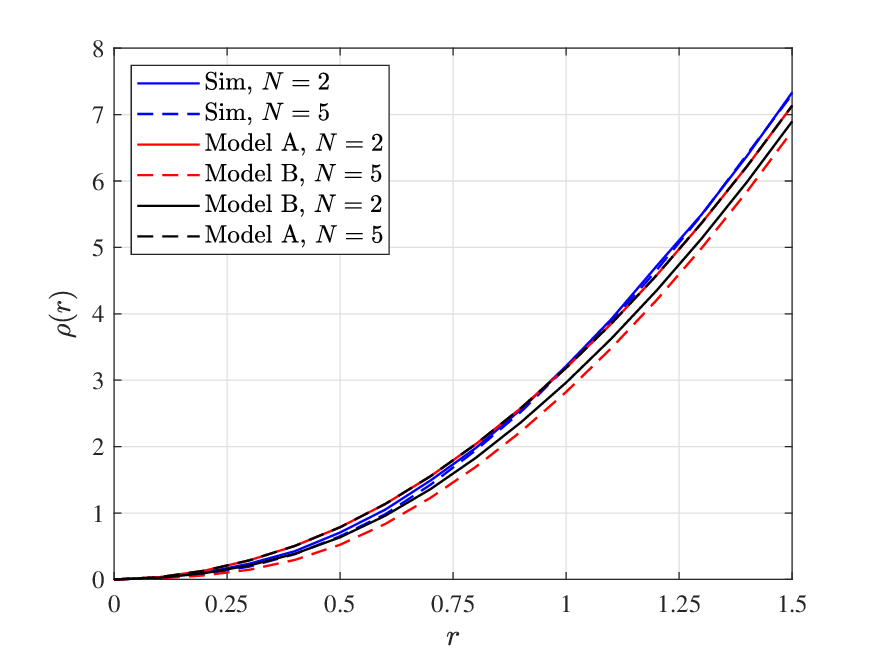}
    \caption{{The simulated second moment measure of $\Phi_{\rm I}$ and the theoretical second moment measures of the candidate approximations ($\Phi_{\rm A}$ and $\Phi_{\rm B}$).}}
    \label{fig_SMM}
\end{figure}

In fact, the intensity measures of these two point processes are identical. It indicates that, from the perspective of the typical BS, the average number of interferers approximated by these two models is consistent~\cite{salehi2018meta}. Moreover, their second moments can be derived as {\small $\mathbb{E}[\bar{N}_{\Phi_{\rm A}}^2(C)]\!=\!N^2\left[\left(\Lambda_{\rm A}^{\rm p} \left(C\right)\right)^2\!+\!\Lambda_{\rm A}^{\rm p}\left(C\right)\right]$} and {\small $\mathbb{E}[\bar{N}_{\Phi_{\rm B}}^2(C)]\!=\!N^2\cdot\left(\Lambda_{\rm A}^{\rm p} \left(C\right)\right)^2\!+\!N \cdot \Lambda_{\rm A}^{\rm p}\left(C\right)$} respectively, where $C\!=\!b(o,r)$ and $\Lambda_{\rm A}^{\rm p} (C)\!=\!\mathbb{E}_o^{!}[\bar{N}_{\Phi_{\rm A}^{\rm p} }(C)]$. It is worth noting that the second moments associated with these two models are identical when $N=1$. We evaluate both models in Fig.~\ref{fig_SMM}, and the PCP $\Phi_{\rm A}$ (model A) provides a better fit to the practical interfering UE point process $\Phi_{\rm I}$ (Simulation) when $N>1$. It more accurately captures the variation in the number of interferers. Hence, it will be adopted in the subsequent analysis. 

%% file: 4-linkLevel.tex
\section{Spatially-Averaged Analysis of Uplink RSMA}
\label{sec:spatial}
In this section, we first derive the Conditional Received Rate (CRR), which represents the received rate for the $n$-th ranked typical UE under a fixed network topology. The CRR is computed using finite MCSs, thereby capturing the discrete rate behavior of practical systems. We then discuss degenerate cases of the CRR, namely NOMA and OMA. Finally, we derive expressions for the spatially-averaged received rate (the first moment of CRR) and  the spatially-averaged achievable rate bounded by the Shannon limit.

\subsection{CRR of uplink RSMA}
For brevity, let $\gamma_{n1}$ and $\gamma_{n2}$ denote 
$\mathrm{SINR}^{\mathrm{RS}}_{n,1}$ and $\mathrm{SINR}^{\mathrm{RS}}_{n,2}$ in~\eqref{eq_SINR_RS}. 
Based on Definition~\ref{def_CRR}, the CRR of the $n$-th ranked typical UE equals  the sum of its sub-message rates, given by
\begin{align}
\label{eq:CRR_RSMA}
&\bar{r}_n^{\text{RS}}(\Phi_{\text{BS}},\Phi_{\rm u})=\mathbb{E}[R_{\text{MCS}}(\gamma_n)|\Phi_{\text{BS}},\Phi_{\rm u}]\notag\\
&=\sum_{m=1}^{M} \!r_m\! \sum_{q=1}^{2}\!\mathbb{P}\left(\theta_m \!< \!\gamma_{nq}<\theta_{m+1}|\Phi_{\text{BS}},\Phi_{\rm u}\right).
\end{align}
The corresponding SE of the $n$-th ranked UE is {\small $\mathrm{SE}_{n}^{\mathrm{RS}}(\Phi_{\mathrm{BS}},\Phi_{\rm u})\!\!=\!\!\bar{r}_n^{\mathrm{RS}}(\Phi_{\mathrm{BS}},\Phi_{\rm u})/BW$}. Substituting~\eqref{eq_SINR_RS} into~\eqref{eq:CRR_RSMA}, we obtain the following lemma.
\begin{lemma}\label{th:CRR_RSMA}
The CRR of the typical BS from its $n$-th ranked typical UE in uplink RSMA transmission is given by
\begin{align}
    \label{eq_CRR_RS_v}
    \bar{r}_n^{\text{RS}}(\Phi_{\text{BS}},\Phi_{\rm u})&\!=\!\sum_{m=1}^{M} \!\left(\Delta r_m \sum_{j=1}^{4}{I_{j,m,n} (\beta)}\right),
\end{align}
where {\small $\Delta r_m\!\!= \!\! r_m\!\! -\!\!r_{m-1}$}, {\small $I_{j,m,n}(\beta)\!\!=\!\!A_{j,m,n}(\beta) f_{j,m,n}(\beta) $}, {\small $A_{j,m,n}(\beta)\!=\!c_j(q) \mathbb{I}(u_{j,m}(\beta)\!>\!0) e^{-\frac{R_n^{\eta}\theta_m \tilde{\sigma}^2}{u_{j,m}(\beta)}}$},
\begin{small}
\begin{align}
    &c_j{(q)}=
    \begin{cases}
        q,& j=1,4,\\
        1-q,& j=2,3,
    \end{cases}
    \notag\\
    &{f_{j,m,n}(\beta)} \!=\! \prod_{i=n+1}^N \!\frac{{u_{j,m}(\beta)}}{{u_{j,m}(\beta)}\!+\!\theta_m\left(R_{n} / R_{i}\right)^{\eta}}\notag\\
    &\times\prod_{x \in \Phi_{\rm I}} \!\frac{{u_{j,m}(\beta)}}{{u_{j,m}(\beta)}\!+\!\theta_m\left(R_{n}/D_x\right)^{\eta}},j\in \{1,2,3,4\},\notag
\end{align}
\end{small}{\small ${u_{1,m}(\beta)}\!=\!(1\!+\!\theta_m)\beta\!-\!\theta_m$}, {\small ${u_{2,m}(\beta)}\!=\!\beta$}, {\small ${u_{3,m}(\beta)}\!=\! 1\!-\!(1\!+\!\theta_m)\beta$}, and {\small ${u_{4,m}(\beta)}\!=\!1\!-\!\beta$}. Note that {\small ${u_{j,m}(\beta)}\!>\!0$} is a sufficient condition for the existence of {\small ${f_{j,m,n}(\beta)}$}, {\small $\forall{j\in \{1,2,3,4\}}$}.
\end{lemma}
\begin{proof}
    see Appendix \ref{app:CRR_RSMA}.
\end{proof}
\begin{remark}
    It is observed that {the CRR} is obtained by two summations: the outer summation depends on the rate discretization strategy, i.e., the choices of rate $r_m$ and threshold $\theta_m$; while the inner summation is determined by the relationship between $\theta_m$ and the allocation fractional factor $\beta$. 
\end{remark}
\begin{corollary}[Special Cases of CRR for NOMA and OMA]
\label{Co:CRR_NOMA}
When {\small $\beta\!=\!0$} or {\small $\beta\!=\!1$}, Lemma \ref{th:CRR_RSMA} reduces to the CRR of the $n$-th {UE} in NOMA transmission
\begin{align}
    \label{eq:rn:NO}
    \bar{r}_n^{\text{NO}}(\Phi_{\text{BS}},\Phi_{\rm u})= \!\sum_{m=1}^{M}\!\Delta r_m {e^{-\theta_m R_{n}^{\eta}\tilde{\sigma}^2}} {f_n(\theta_m)},
\end{align}
where \small ${f_{n}(\theta_m)}\!=\!\prod_{i=n+1}^{N}\!\frac{1}{\theta_{m}R_{n}^{\eta}R_{i}^{-\eta}+1 } \!\prod_{x\in\Phi_{\rm I}}\!\frac{1}{\theta_{m} R_{n}^{\eta}D_{x}^{-\eta}+1}$. If $N=1$, 
the system reduces to {the OMA case}, and the CRR of the sole typical UE is
\begin{align}
    \label{eq:r_oma}
    \bar{r}^{\text{OMA}}(\Phi_{\text{BS}},\!\Phi_{\rm u})\!=\!\sum_{m=1}^{M}\!\!\Delta r_m {e^{\!-\!\theta_m\! R_{1}^{\eta}\tilde{\sigma}^2}}\! \!\!\prod_{x\in\Phi_{\rm I}}\!\!\!\frac{1}{\theta_{m}\! R_{1}^{\eta}D_{x}^{\!-\eta}\!\!+\!\!1}.
\end{align}
\end{corollary}
Unlike RSMA, the CRR expression of NOMA does not include the inner summation, which in RSMA originates from the  fractional power allocation factor $\beta$.

\subsection{Average Received Rate}
Building on Lemma~\ref{th:CRR_RSMA}, we evaluate the spatially-averaged rate performance of uplink RSMA. Specifically, the {average received rate} is defined as the expectation of the CRR over the spatial configurations of $\Phi_{\mathrm{BS}}$ and $\Phi_{\rm u}$.
\begin{theorem}[Average Received Rate for RSMA]
\label{Co:ARRandVar_RSMA}
The spatially-averaged received rate of the $n$-th ranked typical UE in uplink RSMA transmission can be expressed as
\begin{align}
    \label{eq:ARR}
    &\mathbb{E}[\bar{r}_n^{\text{RS}}(\Phi_{\text{BS}},\Phi_{\rm u})]=\sum_{m=1}^{M}\Delta r_m \sum_{j=1}^{4}\mathbb{E}\left[{I_{j,m,n} \left(\beta\right)}\right]\notag\\
    =&2B_1 \pi \lambda (N-n)! \sum_{m=1}^{M}\Delta r_m \sum_{j=1}^{4}\int_{0}^{\infty}  {A_{j,m,n} (\beta)} f_{R_{n}}\left(r\right)\notag\\
   \times & {u_{j,m}\left(\beta\right)}e^{(N-n) B_1 \lambda \pi r^2}{G_{j,m,n}^{\text{RS}}\left(r,\beta\right)T_{j,m}^{\text{RS}}\left(r,\beta\right)}\mathrm{d}r,
\end{align}
where {\small ${G_{j,m,n}^{\text{RS}}(r,\beta)}=\int_{\mathcal{A}}\prod_{i=n+1}^{N}\frac{r_i e^{-B_1 \lambda \pi r_i^2}}{{u_{j,m} (\beta)}+\theta_{m}(r/r_i)^{\eta}}\mathrm{d}\mathbf{r}_n$}, $\mathcal{A}$ denotes the integral area satisfying {\small $r\leq r_{n+1}\leq\cdots \leq r_{N}$}, and {\small $\mathrm{d}\mathbf{r}_n=\mathrm{d}{r_{n+1}}\cdots \mathrm{d}r_{N}$}. {Recall that {\small $B_1=\frac{5}{4}$}}. 
\begin{small}
\begin{align}
    {Q_{j,m}^{\text{RS}}(\!r,\beta,\!x)}\!=\!\!\left[\!1\!\!-\!\left(\!\frac{{u_{j,m}\left(\beta\right)}}{{u_{j,m}\!\left(\beta\right)}\!+\!\theta_m x^{\!-\eta}}\!\right)^{\!N}\right]\!\left(\!1\!-\!e^{\!-B_2 \lambda \pi x^2 r^2}\right)\!x r^2,\notag
\end{align}
\end{small}{where {\small $B_2\!=\!\frac{12}{5}$}} and {\small ${T_{j,m}^{\text{RS}}\left(r,\beta\right)}\!\!=\!\!\exp\left(\!-2\pi \lambda\!\int_{0}^{\infty}\!{Q_{j,m}^{\text{RS}}(r,\beta, x)}\mathrm{d}x\right)$}. 
\end{theorem}
\begin{proof}
    see Appendix \ref{app:ARR_RSMA}.
\end{proof}
Theorem \ref{Co:ARRandVar_RSMA} provides an analytical expression of the average received rate for uplink RSMA transmission. {The} degenerate cases corresponding to NOMA and OMA are omitted here. 
\begin{remark}
\label{Rm:ARR}
The functions {\small $T_{j,m}^{\text{RS}}(r,\beta)$} capture the average effects of inter-cell interference on the $n$-th ranked typical UE. In high path loss environments (e.g., mmWave), inter-cell interference becomes negligible, leading to {\small $T_{j,m}^{\text{RS}}(r,\beta)\to 1$}. Moreover, when $M=1$ and $r_1=1$ (i.e., assigning rate 1 if $\mathrm{SINR} > \theta_1$ and 0 otherwise), the theorem reduces to the  classical coverage probability of SG models~\cite{andrews2011tractable}.
\end{remark}
While the average received rate can provide a rough indication of rate performance, it fails to capture UE-level variations. For example, one UE may achieve 10 Mbps while another is nearly in outage, although the overall average remains high. In Section~\ref{sec:fine}, we will provide fine-grained characterizations of the CRR, including its variability across UEs and reliability.

\subsection{Average Achievable Rate}
To assess the theoretical upper bound of uplink RSMA performance, we consider the Shannon limit under ideal coding and rate adaptation. Specifically, the spatially-averaged achievable rate of the $n$-th ranked UE is defined as\footnote{For clarity, this definition omits the stream-splitting process of RSMA and a refined formulation will be given in the following discussion. Additionally, the inner conditional expectation is taken over other random variables, including channel gains and inter-UE decoding order factors.}
\begin{align}
\bar C_n 
&\triangleq \mathbb{E}_{\Phi_{\mathrm{BS}}, \Phi_u}\Big[\mathbb{E}\big[\ln(1+\mathrm{SINR}_n)|\Phi_{\mathrm{BS}},\Phi_{\rm u}\big]\Big], \label{eq:avg_achievable_rate}
\end{align}
which represents the ergodic rate achievable with Gaussian codebooks and infinitely long blocklengths. By construction, the average received rate with discrete MCS levels in Theorem~\ref{Co:ARRandVar_RSMA} is always bounded above by $\bar C_n$.
\begin{theorem}[Average Achievable Rate]
By continuously adapting to both $\text{SINR}_{n,1}^{\text{RS}}$ and $\text{SINR}_{n,2}^{\text{RS}}$, the average achievable rate for the $n$-th ranked typical UE is given by
\begin{align}
&{\bar C_n^{\text{RS}}}={\mathbb{E}\left[\ln \left(1+\text{SINR}_{n,1}^{\text{RS}}\right)\right]+\mathbb{E}\left[\ln\left(1+\text{SINR}_{n,2}^{\text{RS}}\right)\right]}\notag\\
&=2B_1 \pi \lambda (N-n)!\sum_{j=1}^{4}\int_{0}^{\infty}f_{R_{n}}(r)\int_{0}^{\infty}{A_{j,n}^{\text{SN}}(t,\beta)} \notag\\ 
& \times  L_j (t,\beta) e^{(N-n) B_1 \lambda \pi r^2}G_{n,j}^{\text{SN}}(t,r,\beta)T_{j}^{\text{SN}}(t,r,\beta)\mathrm{d}t \mathrm{d}r,
\end{align}
where 
{\small ${A_{j,n}^{\text{SN}}(t,\beta)}\!\!=\!\!{c_j(q)} \mathbb{I}(L_j (t,\beta)\!\!>\!\!0)e^{-{\frac{r^{\eta}(e^t-1)\tilde{\sigma}^2}{L_j(t,\beta)}}}$}, {\small ${L_1(t,\beta)}\!\!=\!\!e^{t}\beta\!-\!e^t\!+\!1$}, {\small ${L_2(t,\beta)}\!\!=\!\!\beta$}, {\small ${L_3(t,\beta)}\!\!=\!\!1\!-\!\beta e^t$}, and {\small ${L_4(t,\beta)}\!\!=\!\!1\!-\!\beta$}. Additionally, {\small $G_{n,j}^{\text{SN}}(t,r,\beta)\!=\!\int_{\mathcal{A}}\prod_{i=n+1}^{N}\frac{r_i e^{-B_1 \lambda \pi r_i^2}}{L_j (t,\beta) + (e^t-1)(r/r_i)^{\eta}}\mathrm{d}\mathbf{r}_n$}, {\small $T_{j}^{\text{SN}}(t,r,\beta)\!=\!\exp \left(-2\pi\lambda \int_{0}^{\infty}Q_{j}^{\text{SN}}(t,r,\beta,x)\mathrm{d}x\right)$} and 
\begin{small}
\begin{align}
Q_{j}^{\text{SN}}(t,r,\beta,x)\!\!=\!\left[\!1\!-\!\!\left(\frac{L_j(t,\beta)}{L_j(t,\beta)\!+\!\left(\!e^t\!-\!1\right)\!x^{\eta}}\right)^{\!\!N\!}\right]\!\left(\!1\!\!-\!e^{\!-\!B_2 \lambda \pi x^2 r^2}\right)\!\!xr^2\!.\notag
\end{align}
\end{small}
\label{thm:shannon}
\end{theorem}
\begin{proof}
    see Appendix \ref{Pr:AAR}.
\end{proof}
Theorem~\ref{thm:shannon} provides an ideal reference bound under optimal continuous rate adaptation. This result serves as a baseline for quantifying the rate performance gap between practical multi-level MCS strategies and the theoretical limit.

%% file: 5-symLevel.tex
\section{Fine-Grained Analysis of Uplink RSMA}
\label{sec:fine}
In this section, we extend the analysis beyond spatial averages to capture the fine-grained statistical features of uplink RSMA performance. We first derive the moments of the conditional received rate (CRR), including its variance and higher-order statistics, to characterize its distribution. We then introduce the meta distribution to quantify the reliability of the transmission rate across network realizations.

\subsection{Moment of the CRR}
We first present a general expression for the moments of the CRR in uplink RSMA.
\begin{theorem}
\label{th:moments_RSMA}
The $b$-th moment of CRR from the $n$-th ranked typical UE in uplink RSMA is given by
\begin{align}
&M_b^{(n,\text{RS})}=\mathbb{E}\left[\left(\bar{r}_n^{\text{RS}}\left(\Phi_{\text{BS}},\Phi_{\rm u}\right)\right)^b\right]\notag\\
=&\!\!\!\sum_{C_1,C_2}\!\!\!A_b\!\left(\prod_{m=1}^{M}\!\Delta r_{m}^{n_m}\!\right)\!\mathbb{E}\!\!\left[\!\prod_{m=1}^{M}\sum_{C_3,C_4}\!\!\!B_m\!\prod_{j=1}^{4}\!\left({I_{j,m,n}(\beta)}\right)^{k_{j,m}}\!\right]\!,
\end{align}
where {\small $A_b\!=\!\frac{b!}{\prod_{m=1}^{M}n_{m}!}$} {, and the integers {\small $n_m$} satisfy} {\small $C_1\!:\!n_m\!\geq\! 0, 1\!\leq\! m\!\leq M$} and {\small $C_2:\sum_{m=1}^{M}\!n_m\!=\!b$}. {\small $B_m=\frac{n_m!}{\prod_{j=1}^{4}\!k_{j,m}!}$}, {and the integers {\small $k_{j,m}$} satisfy} {\small $C_3:k_{j,m}\geq 0,1\!\leq \!j\!\leq 4$} and {\small $C_4:\sum_{j=1}^{4}\!k_{j,m}\!=\!n_m$}. {\small {$I_{j,m,n}(\beta)$}} is the same as that in {Lemma \ref{th:CRR_RSMA}}.    
\end{theorem}
\begin{proof}
The general analytical form of the CRR moments is obtained by applying a double multinomial series expansion.
\end{proof}
\begin{remark}
The variance of the CRR can be obtained from Theorem~\ref{th:moments_RSMA} as 
{\small $M_2^{(n,\mathrm{RS})}- \big(M_1^{(n,\mathrm{RS})}\big)^2$}.
Higher-order statistics (e.g., skewness and kurtosis) can be obtained in this way. Under the normalized bandwidth assumption ({\small $BW\!=\!1$}), the spatially-averaged SE and the variance of the conditional SE are equivalent to those of the CRR.
\end{remark}
\begin{corollary}[NOMA case]
\label{Co:moments_NOMA}
    In uplink NOMA, the $b$-th moment of the CRR for the $n$-th ranked associated UE is given by 
    \begin{align}
        &M_{b}^{(n,\text{NO})}\!=\!2 B_1 \pi \lambda (N\!-\!n)! \sum_{C_1,C_2}
        A_b \left(\prod_{m=1}^{M} \Delta r_{m}^{n_{m}}\right)\notag\\
        & \times \int_{0}^{\infty}f_{R_{n}}\left(r\right) e^{K_n(r)}G_n(r)T(r)\mathrm{d}r,
    \end{align}
    where {\small $K_n(r)\!\!\!=\!\!\!(N\!-\!n)B_1 \lambda\pi r^2\!\!-\!\!r^{\eta}\tilde{\sigma}^2 \Omega$}, {\small $\Omega\!\!=\!\!\sum_{m=1}^{M}\theta_m n_m$}, {\small $G_n(r)\!\!=\!\!\!\int_{\mathcal{A}}\!\prod_{m=1}^{M}\prod_{i=n+1}^{N}\frac{r_i e^{\!-B_1 \lambda \pi r_i^2}}{\left[1\!+\!\theta_m \left(r/r_i\right)^{\eta}\right]^{n_m}}\mathrm{d}\mathbf{r}_n$}, 
{\small $T(r)\!\!=\!\exp\left(\!-\!2\pi\lambda\! \int_{0}^{\infty}\!\!Q(r,x)\mathrm{d}x\right)$} and 
\begin{small}
\begin{align}
    Q(r,\!x)\!=\!\left[1\!\!-\!\!\!\prod_{m=1}^{M}\!\!\left(\frac{1}{\theta_m x^{-\!\eta}\!+\!1}\right)^{\!N\cdot n_m}\right]\!\left(\!1\!\!-\!e^{\!-\!B_2\lambda \pi x^2 r^2}\right)\!xr^2.\!\notag
\end{align}
\end{small}
\end{corollary}
\begin{proof}
    see Appendix \ref{Pr:Co4}.
\end{proof}
\begin{remark}
When $M=1$ and $r_1=1$, the $b$-th moment of the CRR reduces to that of the conditional success probability in uplink NOMA without considering MCS, which corresponds to the Theorem $1$ in~\cite{salehi2018meta}.
\end{remark} 
\begin{corollary} [OMA case]
\label{Co:noma-oma}
    When $N=1$, i.e., the NOMA system degenerates into an OMA system, the $b$-th moment of the CRR is given by
    \begin{align}
        &M_{b}^{\text{OMA}} \!=\!\sum_{C_1,C_2}\!A_b\! \left(\prod_{m=1}^{M} \Delta r_{m}^{n_{m}}\right)\!\int_{0}^{\infty}\!f_{R_1}\left(r\right){e^{-r^{\eta}\tilde{\sigma}^2 \Omega}}T(r)\mathrm{d}r, 
    \end{align}  
    where $T(r)$ follows from the one in Corollary \ref{Co:moments_NOMA}, and $C_1$ and $C_2$ are the same as those in Theorem~\ref{th:moments_RSMA}. 
\end{corollary}
\begin{remark} 
    For the $N=1$ case, the intra-cell interference vanishes and the inter-cell interference reduces to an inhomogeneous PPP with intensity $\lambda_{\rm I}(r)$ as shown in Section~\ref{sec:inter}. This case corresponds to {the} Theorem 1 in~\cite{guo2025stochastic} when there is no power control and full user activity.
\end{remark}

\subsection{Meta Distribution of CRR}
\label{section:meta}
The meta distribution of the CRR provides the information about the fraction of UEs whose rates exceed a given threshold, or the probability distribution of a typical link's rate.~\cite{wang2017meta}.
\begin{definition}[Meta Distribution of CRR for Uplink RSMA]
Given a data rate threshold $\xi$ and a set of SINR thresholds $\Theta$, the meta distribution of CRR of the $n$-th ranked typical UE in uplink RSMA transmission is defined as
\begin{align}
    \label{eq:meta}
    \bar{F}_n\left(\Theta,\xi\right)=\mathbb{P}[\bar{r}_n (\Phi_{\text{BS}},\Phi_u)>\xi],
\end{align}
where $\xi\in \mathbb{R}^{+}$. For instance, if $\bar{F}=0.9$ and $\xi=2$, it means that 90\% UEs achieves at least $2$. Or equivalently, a typical UE achieves $2$ with probability $0.9$. This definition applies to all transmission schemes considered, so we omit superscripts such as ``RS'', ``NO'', and ``OMA'' here.
\end{definition}
The meta distribution of the $n$-th typical UE’s CRR can be obtained by applying the Gil-Pelaez inversion theorem~\cite{haenggi2015meta}
\begin{align}
    \bar{F}_n(\Theta,\xi)=\frac{1}{2}+\frac{1}{\pi}\int_{0}^{\infty}\frac{1}{w}\Im\{u^{-iw}M_b^{(n)}(iw)\}\mathrm{d}w,
\end{align}
where $i=\sqrt{-1}$, $\Im$ denotes the the imaginary part. Recall that $M_b^{(n)}(\cdot)$ represents the moment function{, }valid across different transmission schemes. Since the exact meta distribution has a prohibitive computational cost {and the CRR distribution follows a monotonic curve intuitively}~\cite{wang2019meta}, we adopt the beta distribution as a tractable approximation for $[0,1]$-supported random variables. It indicates that we should normalize the CRR variable {with its range to be} the unit interval before approximation. Recall that for a random variable $X$ following a beta distribution, its PDF is {\small $g(x;\alpha,\phi)\!=\!\frac{1}{B(\alpha,\phi)}x^{\alpha-1}(1\!-\!x)^{\phi-\!1}$}, where {\small $B(\alpha,\phi)\!=\!\frac{\Gamma(\alpha)\Gamma(\phi)}{\Gamma(\alpha\!+\!\phi)}$} denotes beta function. $\alpha$ and $\phi$ determine the shape of the distribution, which can be calculated by $\mathbb{E}[X]\!=\!\frac{\alpha}{\alpha+\phi}$ and $\mathrm{Var}[X]\!=\!\frac{\alpha\phi}{(\alpha+\phi)^2 (\alpha+\phi+1)}$. Thus, we only need to acquire the first- and second-order statistical properties of CRR, derived in Theorem~\ref{th:moments_RSMA}, to estimate its meta distribution, i.e., $\mathbb{P}[\bar{r}_n (\Phi_{\text{BS}},\Phi_u)>\xi]$, via the beta distribution.

%% file: 6-sim.tex
\section{Numerical Results}
\label{sec:sim}
In this section, we validate the {derived} analytical results through Monte Carlo simulations. We first evaluate the spatially averaged received and achievable rates, then assess the CRR variance and the CRR meta distribution. Concurrently, we provide a comparative analysis of {RSMA and NOMA} schemes to quantify the performance gaps. In addition, we analyze the impacts of key system parameters, including MCS configuration, noise power, BS intensity, inter-UE decoding order factor, and path loss exponent.  

Consider an uplink cellular network where BSs are distributed according to a PPP with density $\lambda_{\mathrm{BS}}$. The UE density is assumed to be sufficiently high so that each cell contains at least $N$ {UEs}. Unless otherwise specified, the network operates in the interference-limited regime ($\sigma^2=0$) with a path-loss exponent $\eta=4$, and the {inter-UE} decoding order is fixed (i.e., $q\!=0$). Bandwidth is normalized to {$BW=1$}. 
The simulation procedure is as follows:
\begin{itemize}
  \item[1)] \textit{Network generation:}  
  A square area of \(1\,\mathrm{km} \times 1\,\mathrm{km}\) is considered {and {\small $\lambda_{\mathrm{BS}} = 10^{-4}$}}. UEs are uniformly distributed within each BS Voronoi cell. 
  Each BS serves $N=\!2\!$ UEs for {RSMA or NOMA}. 
  A total of $N_{\mathrm{iter}} = 10^{4}$ independent spatial realizations are simulated.

  \item[2)] \textit{Channel realization:}  
  For each topology, {\(N_{\mathrm{iter}}^{1}=5\times 10^{4}\)} independent small-scale fading samples are generated 
  to account for channel randomness.

  \item[3)] \textit{SINR computation:}  
  For each realization, the SINR values of the $n$-th ranked typical UE 
  are computed according to {\eqref{eq_SINR_RS} and \eqref{eq_SINR_NO}}, 
  depending on the access scheme.

  \item[4)] \textit{Rate evaluation:}  
  The instantaneous achievable rate is obtained as 
  \(\ln(1+\mathrm{SINR})\),  
  while the received rate with MCS adaptation is computed as Definition~\ref{def_CRR}, 
  determined by the SINR thresholds \(\Theta\) and discrete rates $\{r_{i_{\mathrm{CQI}}}\}_{0}^{M}$.

  \item[5)] \textit{Statistical aggregation:}  
  By averaging the simulated rates over channel realizations and spatial topologies, 
  we estimate the spatially-averaged received rate, achievable rate, and the variance of the CRR.  
  The proportion of CRR values exceeding a threshold~\(\xi\) approximates the CRR meta distribution.
\end{itemize}

The MCS adaptation settings are designed to illustrate the qualitative impact of 
different granularities of adaptation on uplink rate performance. 
The SINR thresholds and discrete rates follow a simplified quasi-logarithmic mapping, 
capturing the general trend of adaptive modulation and coding systems 
while keeping the analysis analytically tractable. 
In this setting, four representative adaptation strategies are considered:  
\texttt{S1}: \(\theta_1\!=\!-15\,\mathrm{dB}\); 
\texttt{S2}: \(\theta_1\!=\!-5\,\mathrm{dB}\); 
\texttt{S3}: \(\Theta\!=\!\{-15,-10\}\,\mathrm{dB}\); 
\texttt{S4}: \(\Theta\!=\!\{-15,-5\}\,\mathrm{dB}\). Unless otherwise stated, the discrete rates are \(r_1=0.4\) and \(r_2=0.6\), 
which represent normalized rates achievable under different modulation–coding combinations. This abstraction focuses on the qualitative influence of MCS granularity rather than on any specific implementation, providing a generic setting for performance evaluation.

\subsection{Spatially-averaged Rate}
\begin{figure}[t]
    \centering
    \begin{minipage}{0.4\textwidth}
        \centering
        \includegraphics[width=\linewidth]{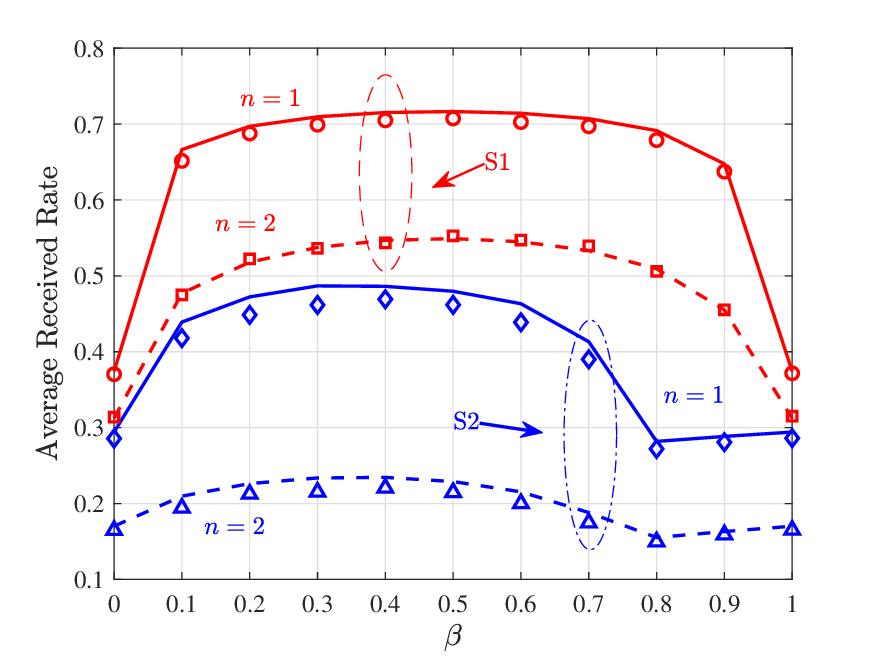} 
        \caption*{{(a) \texttt{S1} and \texttt{S2}}}
    \end{minipage}
    \begin{minipage}{0.4\textwidth}
        \centering
        \includegraphics[width=\linewidth]{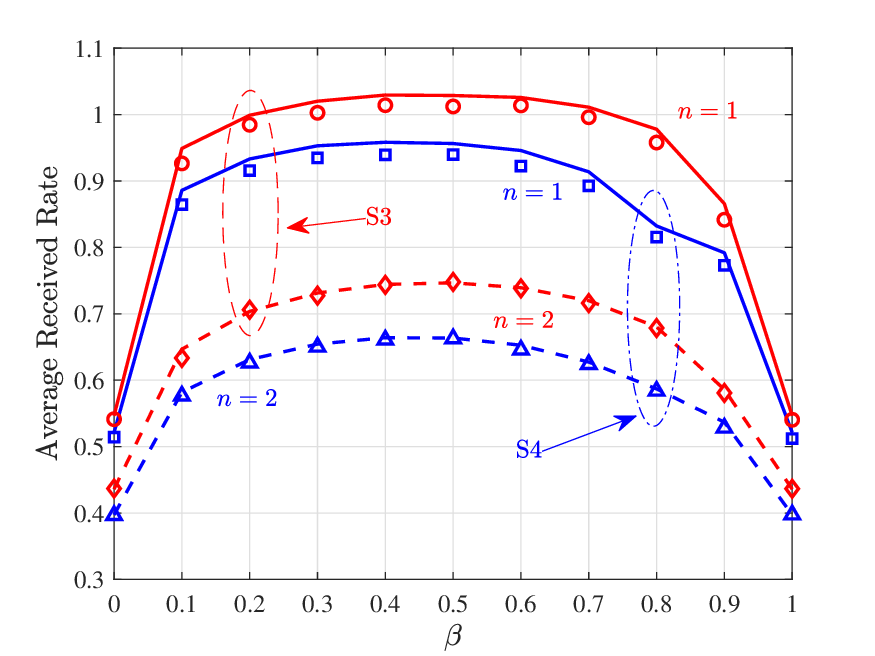}
        \caption*{{(b) \texttt{S1} and \texttt{S2}}}
    \end{minipage}
    \caption{{The average received rate with $N=2$ selected UEs per cell. $n=1$ and $n=2$ denote the near and far UE, respectively. Curves denote analytical results in Theorem~\ref{Co:ARRandVar_RSMA}, and markers denote simulation results.}}
    \label{fig_RSMA_N2_q0}
\end{figure}

Fig.~\ref{fig_RSMA_N2_q0} (a) shows the spatially-averaged received rate versus the rate-splitting factor~$\beta$ for the single-level MCS case ($M=1$). Both analytical and simulation results are provided, and their close agreement confirms the accuracy of Theorem~\ref{Co:ARRandVar_RSMA}. Two SINR thresholds, \(\theta_1=-15\) dB (\texttt{S1}) and \(\theta_1=-5\) dB (\texttt{S2}), are considered for comparison. As $\beta$ varies, both UEs exhibit a quasi-concave rate profile, indicating the existence of an optimal splitting factor~$\beta^\star$ that maximizes the average received rate. The near UE ($n=1$) consistently achieves a higher rate than the far UE ($n=2$) owing to its stronger link quality, while both UEs benefit noticeably from rate splitting. This behavior confirms that a power allocation ($0\!<\beta^\star\!<1$) can effectively balance the interference and decoding complexity, outperforming the extreme cases of pure NOMA ($\beta\in\{0,1\}$). Furthermore, increasing the SINR threshold from \(-15\) dB to \(-5\) dB results in a performance degradation 
because the probability that the instantaneous SINR exceeds the threshold becomes smaller. 
Fig.~\ref{fig_RSMA_N2_q0} (b) shows the spatially-averaged received rate as a function of~$\beta$ 
for the two-level MCS configurations (\(M=2\), corresponding to \texttt{S3} and \texttt{S4}). Compared with the single-level case in Fig.~\ref{fig_RSMA_N2_q0} (a), introducing an additional MCS level increases the overall rate, as the finer quantization of achievable rates allows better adaptation to channel variations. Both UEs exhibit {quasi-concave} rate profiles with respect to~$\beta$, and the optimal splitting factors remain close to those in the $M=1$ case, indicating that RSMA maintains its power–interference balance under higher-order adaptation. Between the two configurations, \texttt{S3} achieves slightly higher rates than \texttt{S4}, since its smaller second threshold enables a higher proportion of UEs to access the upper-rate MCS level. This result illustrates how the placement of MCS thresholds affects system performance and confirms that finer adaptation granularity consistently improves the spatially-averaged rate.

\begin{figure}[t]
    \centering
    \begin{minipage}{0.4\textwidth}
        \centering
        \includegraphics[width=\linewidth]{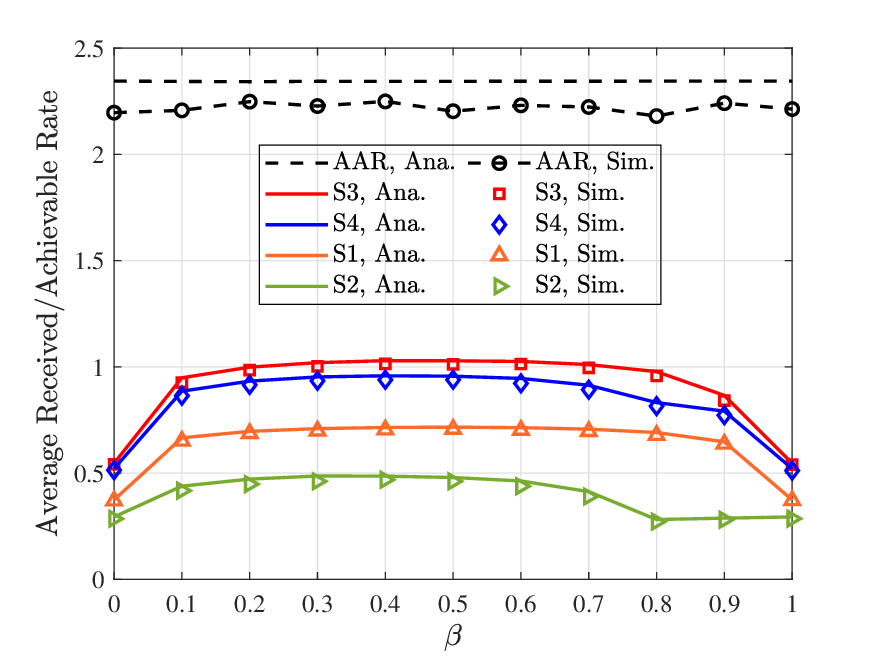} 
        \caption*{{(a) The near/first-ranked typical UE}}
    \end{minipage}
    \begin{minipage}{0.4\textwidth}
        \centering
        \includegraphics[width=\linewidth]{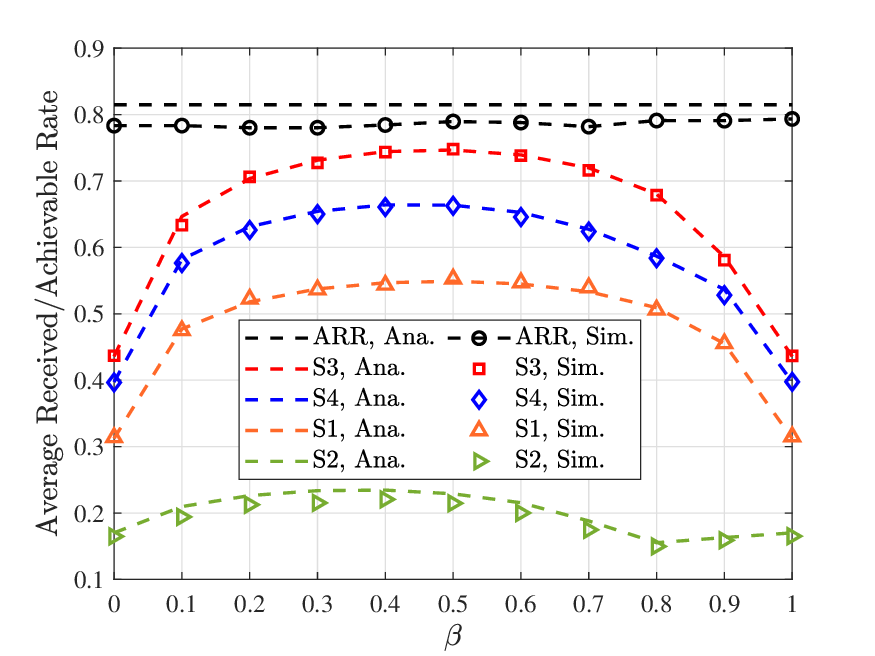}
        \caption*{{(b) The far/second-ranked typical UE}}
    \end{minipage}
    \caption{{Comparison of the average achievable rate and the average received rate of \texttt{S1}, \texttt{S2}, \texttt{S3}, and \texttt{S4} with $N=2$ UEs per cell. The analytical curve of the average achievable rate (as a function of $\beta$) is plotted based on Theorem~\ref{thm:shannon}.}}
    \label{fig_RSMA_N2_vs_q0}
\end{figure}

Fig.~\ref{fig_RSMA_N2_vs_q0} compares the average received rate under discrete MCS adaptation with the average achievable rate obtained from continuous SINR adaptation (Shannon upper bound). The endpoints at $\beta\!\in\!\{0,1\}$ correspond to the NOMA configurations. As expected, the upper bound is essentially insensitive to the rate splitting factor $\beta$, whereas the average received rate follows a {quasi-concave} trend that peaks at moderate $\beta$ values. Under the current configuration, where both UEs adopt the same MCS parameters,  the far user ($n\!=\!2$) nearly attains the bound around $\beta\!\approx\!0.5$, while for the near UE ($n\!=\!1$) a noticeable gap persists over the entire range of $\beta$. This performance difference likely reflects the quantization effect of discrete MCS adaptation, which tends to favor UEs operating in low-SINR regimes under the present settings. In contrast, high-SINR UEs cannot fully exploit their SINR margin due to the limited granularity of discrete rate levels. Comparing the NOMA and RSMA cases, we observe that incorporating the rate splitting principle consistently improves average received rates, confirming RSMA's robustness under practical MCS adaptation.

\subsection{Variance of CRR}
\begin{figure}[t]
    \centering
    \includegraphics[width=0.4\textwidth]{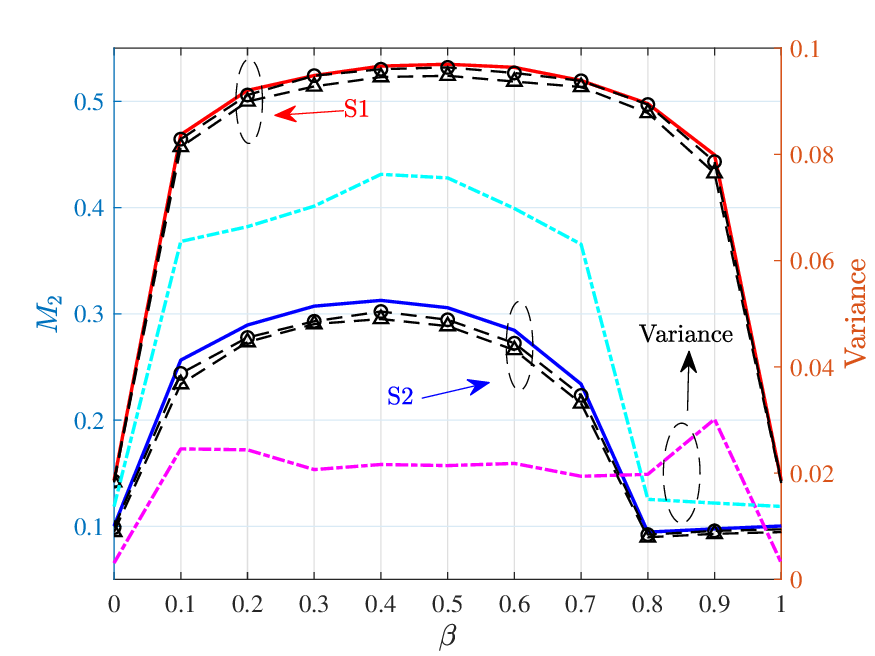}
    \caption{{Second moment $M_2$ (left axis) and variance (right axis) of the first typical UE vs. $\beta$ under \texttt{S1} and \texttt{S2}. Solid: analytical; dashed with markers: simulations with/without co-location.}}
    \label{fig_RSMA_N2_M2_Var_q0}
\end{figure}

Fig.~\ref{fig_RSMA_N2_M2_Var_q0} shows the second moment ($M_2$) and the corresponding variance of the CRR for the first typical UE as functions of $\beta$, under the single-threshold settings \texttt{S1} ($\theta_1=-15$\,dB) and \texttt{S2}  ($\theta_1=-5$\,dB). 

The left axis presents $M_2$, which confirms Theorem~\ref{th:moments_RSMA}. 
Analytical curves are compared with simulation results obtained with and without the co-location.   
Specifically, the co-location case corresponds to \textit{Model~I} (Poisson cluster process) in Section~\ref{sec:inter}, where the interference is computed by placing the $N$ intra-cell UEs at a common location.
The non–colocated case represents a more realistic setting where UEs are uniformly distributed within each cell.    
Both simulation settings closely match the analytical curves, confirming the validity of the PCP assumption and the derived expressions.  Furthermore, increasing $\theta_1$ leads to smaller $M_2$ values over the entire range of  $\beta$. For each $\theta_1$, $M_2$ exhibits a minimum and a maximum at distinct $\beta$ values.

The right axis depicts the CRR variance derived from the first and second moments.  The light blue and magenta dashed lines represent the results for \texttt{S2} and \texttt{S1}, respectively. 
Across most $\beta$ values, a larger $\theta_1$ (i.e., {light blue one}) yields higher variance, indicating greater fluctuation in rate performance among users. Note that an optimal $\beta$ can be identified to minimize CRR variance, thereby improving UE fairness in rate performance.

\subsection{Meta Distribution of CRR}
\begin{figure}[t]
    \centering
    \begin{minipage}{0.4\textwidth}
        \centering
        \includegraphics[width=\linewidth]{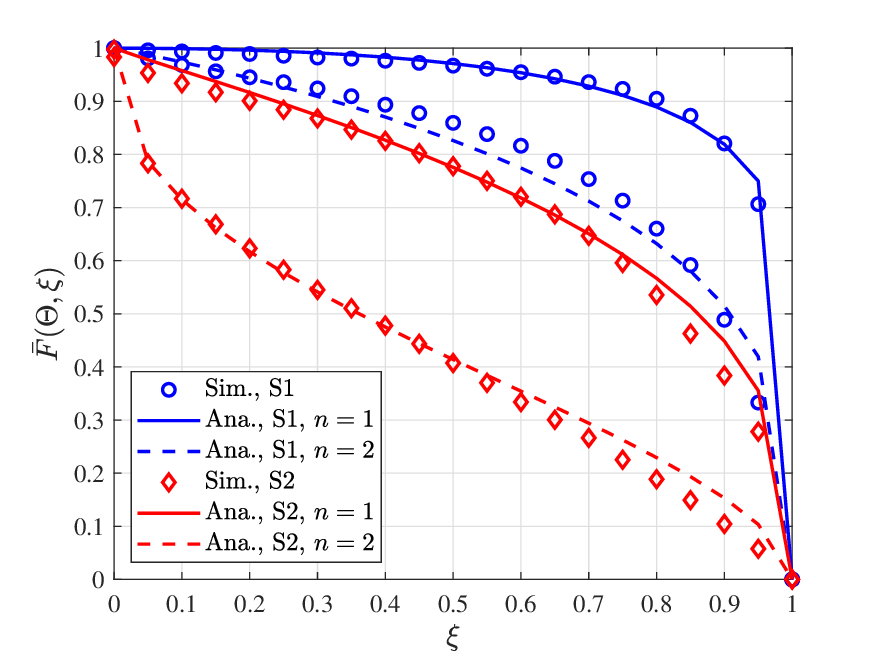} 
        \caption*{{(a) NOMA ($\beta=0$)}}
    \end{minipage}
    \begin{minipage}{0.4\textwidth}
        \centering
        \includegraphics[width=\linewidth]{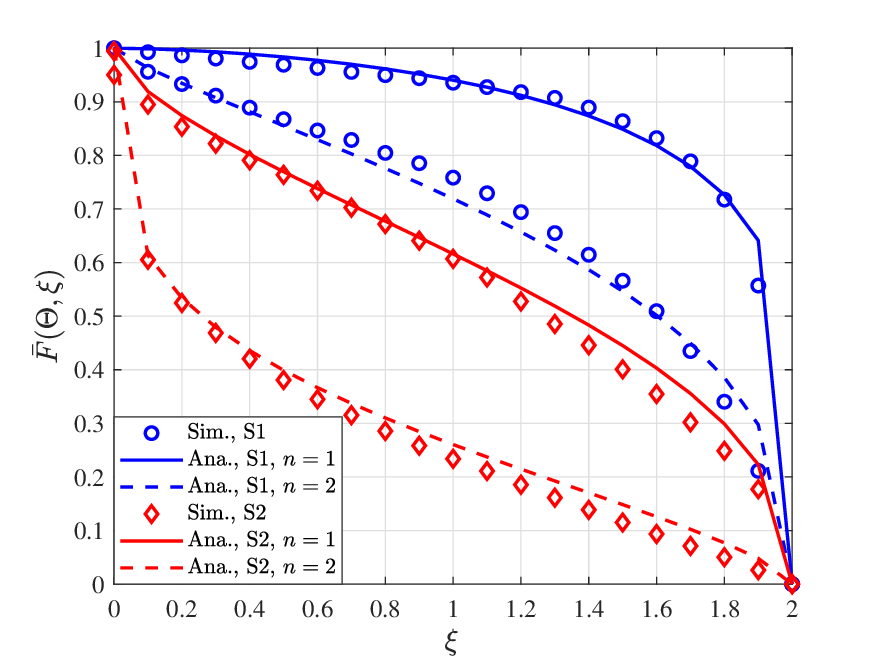}
        \caption*{{(b) RSMA ($\beta=0.5$)}}
    \end{minipage}
    \caption{{The meta distribution of CRR vs. the given rate threshold $\xi$ for $N=2$ under \texttt{S1} and \texttt{S2}.}}
    \label{fig_Meta_q0_N2}
\end{figure}

Fig.~\ref{fig_Meta_q0_N2} (a) depicts the meta distribution of the CRR for $\beta=0$ and $r_1=1$, corresponding to the NOMA case. 
For $\xi=0.2$, the complementary CDF $\bar{F}(\Theta,\xi)$ indicates that the first and second UEs achieve rates above $0.2$ with probabilities of approximately 0.9 and 0.6, respectively. This implies that 90\% of the first UEs and 60\% of the second UEs can sustain a rate higher than $0.2$. Furthermore, the performance gap between the two UEs is more significant in scenario~\texttt{S2} than in \texttt{S1}, since the SINR values of UE~2 more frequently fall within the range $[-15, -5]$~dB than UE~1, leading to higher rate variability. Fig.~\ref{fig_Meta_q0_N2} (b) shows the meta distribution of the CRR for $\beta=0.5$, corresponding to the equal power-splitting case of RSMA. 
Since the beta distribution approximates the meta distribution, the CRR values are first normalized to the unit interval $[0,1]$ before curve fitting and then rescaled back to the rate domain for visualization consistency. Compared with the NOMA case ($\beta=0$), rate splitting effectively enlarges the rate region {from {\small $[0,1]$} to {\small $[0,2]$}}. This indicates that typical UEs have a non-zero probability of achieving transmission rates exceeding 1, or that the percentage of UEs achieving a rate higher than 1 is non-zero.
\begin{figure}[t]
    \centering
    \includegraphics[width=0.4\textwidth]{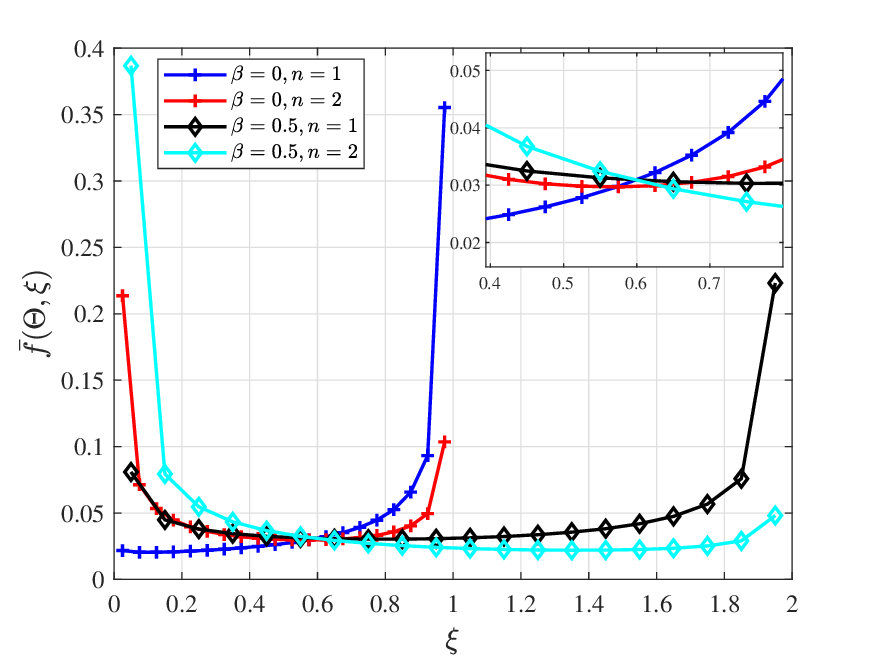}
    \caption{{The interval probabilities for meta distribution of CRR vs. $\xi$ with \texttt{S2}. The results are calculated according to the analytical outcomes in Fig.~\ref{fig_Meta_q0_N2}}.}
    \label{fig_df_Meta}
\end{figure}
Fig.~\ref{fig_df_Meta} illustrates the interval probabilities of the meta distribution of the CRR, 
defined as {\small $\bar{f}(\Theta,\xi) = \bar{F}(\Theta,\xi-\Delta)-\bar{F}(\Theta,\xi)$}, where $\Delta=0.05$. This metric quantifies the probability that the conditional received rate $\bar{r}_n(\Phi_{\mathrm{BS}}, \Phi_u)$ lies within $[\xi-\Delta,\,\xi)$, reflecting how UE rates are distributed across different rate levels. Compared with NOMA ($\beta=0$), RSMA ($\beta=0.5$) reshapes the rate distribution to a more dispersed form. In particular, UE rates originally concentrated in medium-rate regions partly shift toward higher and slightly lower rate levels when RSMA is applied. On average, RSMA yields higher rates than NOMA, indicating an overall {rate improvement}.

\subsection{Impact of Parameters on Average Received Rates}
\begin{figure}[t]
    \centering
    \begin{minipage}{0.4\textwidth}
        \centering
        \includegraphics[width=\linewidth]{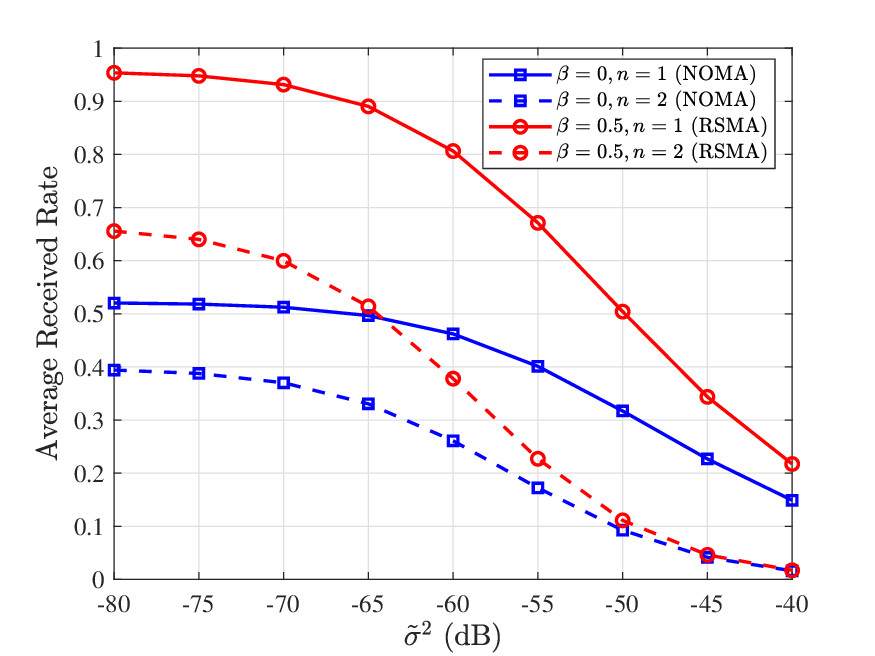} 
        \caption*{{(a) The average received rate vs. noise power $\tilde{\sigma}^2$}.}
    \end{minipage}
    \begin{minipage}{0.4\textwidth}
        \centering
        \includegraphics[width=\linewidth]{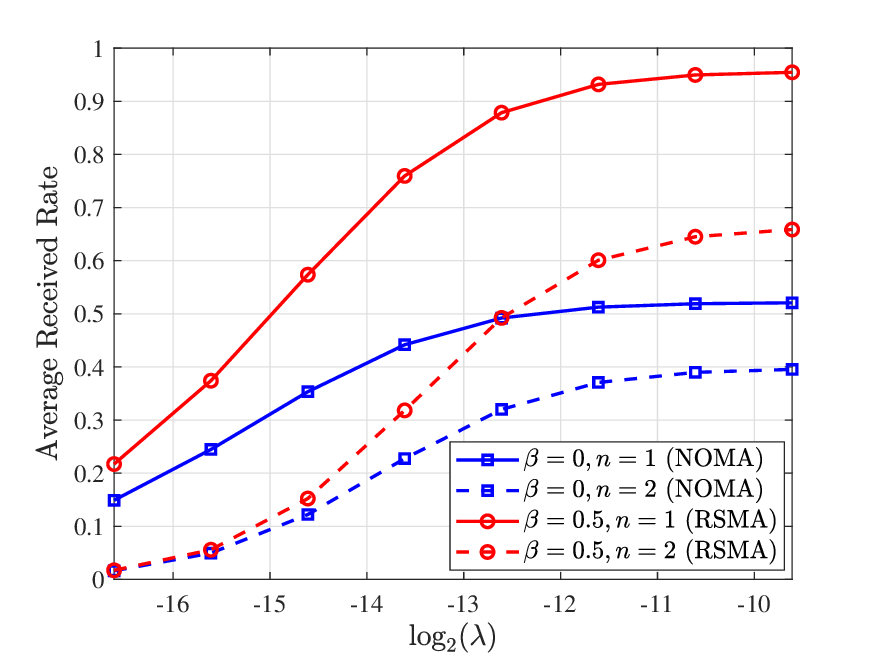}
        \caption*{{(b) The average received rate vs. BS intensity $\lambda$.}}
    \end{minipage}
    \caption{{The impact of $\tilde{\sigma}^2$ and $\lambda$ on the average received rate with \texttt{S4} in NOMA and RSMA systems. }}
    \label{fig_RSMA_N2_M2_sigma_lambda_im}
\end{figure}

In this part, we investigate the effects of the normalized noise power $\tilde{\sigma}^2$, BS density $\lambda$, inter-UE decoding order factor $q$, and path loss exponent $\eta$ on the average received rate. Fig.~\ref{fig_RSMA_N2_M2_sigma_lambda_im} (a) shows the impact of the normalized noise power $\tilde{\sigma}^2$ on the average received rate. Overall, the average received rate decreases as $\tilde{\sigma}^2$ increases for both UEs. In the high-noise regime, the average rate drops significantly because noise dominates the signal power. In the low-noise regime, the average rate exhibits a saturation behavior as $\tilde{\sigma}^2$ decreases. The gain from rate splitting approaches zero and a positive constant in the high- and low-noise regimes, respectively. It indicates that RSMA mainly benefits interference-limited rather than noise-limited scenarios. Fig.~\ref{fig_RSMA_N2_M2_sigma_lambda_im} (b) shows how the BS density~$\lambda$ influences the average received rate under a noisy condition\footnote{In the absence of noise, the average received rate remains invariant to~$\lambda$ because both the desired signal and the interference scale homogeneously with the BS–UE distance.}.  For both UEs, the average received rate increases with $\lambda$ and gradually plateaus at higher densities. 
The joint impact of a higher BS density can explain this trend. As $\lambda$ grows, both the desired signal and the aggregate interference become stronger due to shorter BS–UE distances and more active transmitters. At low and moderate densities, the signal power gain dominates, leading to an overall increase in SINR. 
When the network becomes dense enough, interference dominates and noise becomes negligible, driving the system toward an interference-limited regime where both signal and interference scale homogeneously, hence the observed plateau. Overall, RSMA achieves consistently higher average rates than NOMA across~$\lambda$.

\begin{figure}[t]
    \centering
    \includegraphics[width=0.4\textwidth]{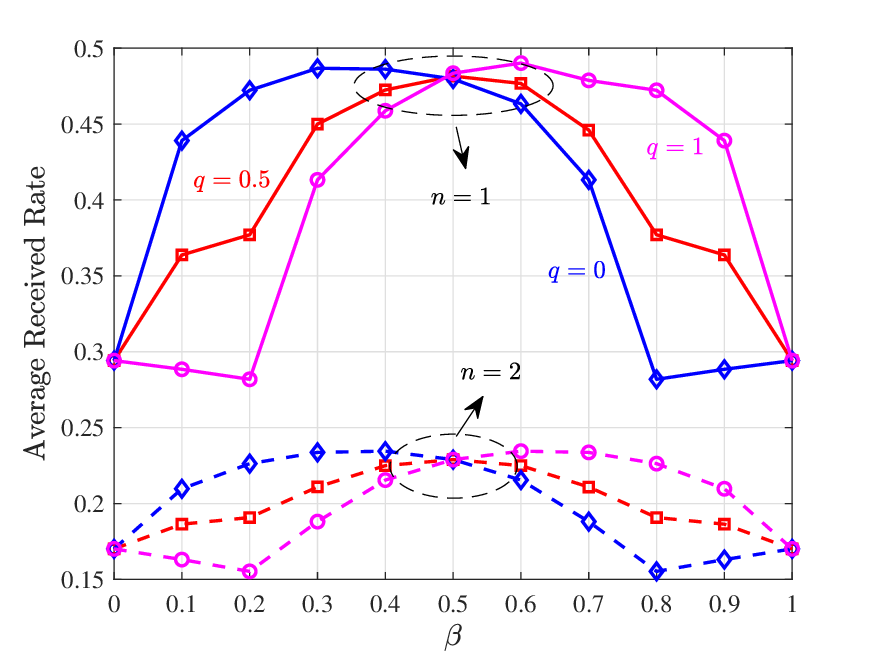}
     \caption{{The average received rate vs. $\beta$ across different inter-UE decoding order factor $q$ under \texttt{S2}. $q=0$, $ q=0.5$, and $ q=1$ represent UE $2$ first, random, UE $1$ first decoding strategies, respectively.}}
    \label{fig_RSMA_N2_M1_q_im}
\end{figure}

Fig.~\ref{fig_RSMA_N2_M1_q_im} illustrates how the average received rate varies with $\beta$ for different values of the inter-UE decoding order factor $q$. Three representative cases are considered: $q=0$, $ q=0.5$, and $ q=1$. When $q=0$ or $q=1$,  the BS always prioritizes the decoding of sub-messages indexed by 2 or 1, respectively. This implies that the random variables $b_n$ degenerate to constants $0$ or $1$. In contrast, $q=0.5$ corresponds to equiprobable decoding orders, where the two sub-messages contribute equally to the average rate. Furthermore, the curves obtained for $q=0$ and $q=1$ also exhibit symmetry with respect to $\beta=0.5$. This is because the decoding priorities of the two sub-messages are interchanged between these two cases.

\begin{figure}[t]
    \centering
    \includegraphics[width=0.4\textwidth]{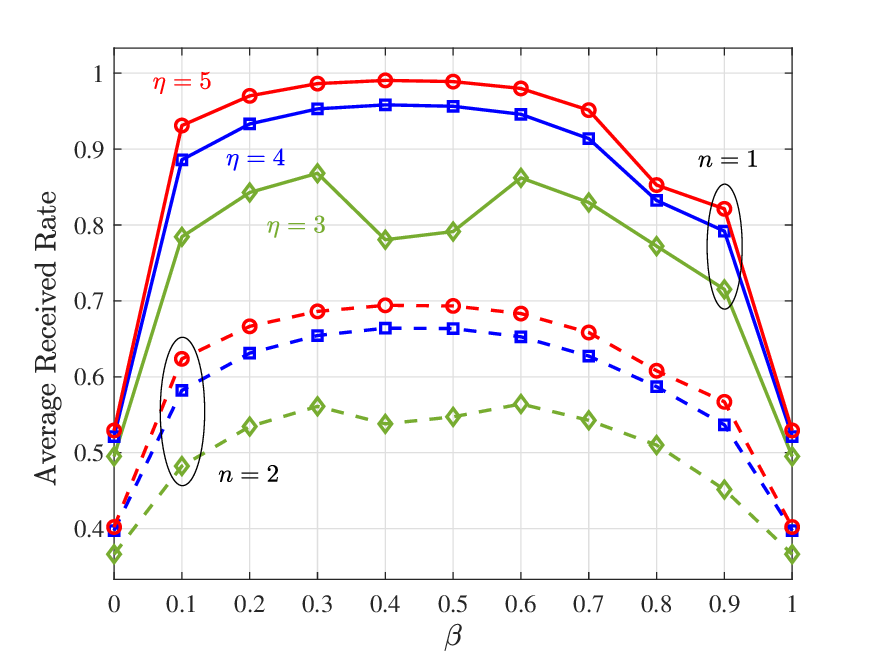}
    \caption{{The average received rate vs. $\beta$ under \texttt{S4} across different pass-loss exponent $\eta$.}}
    \label{fig_RSMA_N2_M2_alpha_im}
\end{figure}

Fig.~\ref{fig_RSMA_N2_M2_alpha_im} shows the average received rate for three path loss exponents, $\eta \in \{3,4,5\}$. A clear increasing trend is evident: a larger $\eta$ leads to a higher average received rate. This is because interference decreases faster than the desired signal power when $\eta$ increases, which improves the overall SINR. As a result, both users benefit from better link conditions. In addition, changes in $\eta$ slightly modify the shape of the curves, indicating that the system performance is sensitive to propagation conditions.

%% file: 7-conclusion.tex
\section{Conclusion}
\label{sec:conclusion}
This paper presented a unified analytical framework for uplink RSMA networks, integrating stochastic geometry with discrete MCS-based rate adaptation to bridge theoretical tractability and practical transmission realism. The framework captured the spatially coupled interference inherent in multi-cell topologies and quantifies performance through the CRR and its key statistics—spatial averages, moments, and meta distributions. These results provided a comprehensive view of rate behavior, revealing how discrete rate adaptation and rate-splitting jointly influence transmission rate, and its variability and reliability. Beyond RSMA, the proposed model generalized naturally to classical multiple access schemes such as NOMA and OMA, thereby offering a common analytical ground for fair comparison across access paradigms. Future work will extend this framework to include advanced network features such as user clustering, and optimization-driven design, further enhancing its applicability to 6G and beyond network analysis.

%% file: appendix.tex
\appendix
\subsection{Proof of Lemma $1$}
\label{app:CRR_RSMA}
Given the point processes of BSs and UEs, the CRR of the typical BS from the $n$-th ranked typical UE with the RSMA transmission scheme is given by
\begin{align}
&\bar{r}_n(\Phi_{\text{BS}},\Phi_{\rm u})=\mathbb{E}[R_{\text{MCS}}(\gamma_n)|\Phi_{\text{BS}},\Phi_{\rm u}]\notag\\
&=\sum_{m=1}^{M} r_m \sum_{q=1}^{2}\mathbb{P}\left(\theta_m \!< \!\gamma_{nq}<\theta_{m+1}|\Phi_{\text{BS}},\Phi_{\rm u}\right)\notag\\
&=\sum_{m=1}^{M}\Delta r_m \sum_{q=1}^{2}\mathbb{P}\left(\gamma_{nq}>\theta_m |\Phi_{\text{BS}},\Phi_{\rm u}\right),
\end{align}
where $\Delta r_m = r_m-r_{m-1}$. Given the similarity in the proofs of $\mathbb{P}(\gamma_{n1}>\theta_m|\Phi_{\text{BS}},\Phi_{\rm u})$ and $\mathbb{P}(\gamma_{n2}>\theta_m|\Phi_{\text{BS}},\Phi_{\rm u})$, we shall focus on deriving $\mathbb{P}(\gamma_{n1}>\theta_m|\Phi_{\text{BS}},\Phi_{\rm u})$ in what follows. The conditional probability of $\gamma_{n1}$ for the $n$-th ranked typical UE is given by
\begin{align}
\label{eq:gamma1}
&\mathbb{P}\left(\gamma_{n1}>\theta_m |\Phi_{\text{BS}},\Phi_{\rm u}\right)\notag\\
=&q\!\times\!\underbrace{\mathbb{P}\left(\!\frac{\beta H_{n}^{o} R_{n}^{-\eta}}{H_{n}^{o} R_{n}^{-\eta}\left(\!1\!-\!\beta\right)\!+\!I_{n}^{\text{intra}}\!+\!I^{\text{inter}}\!+\!\tilde{\sigma}^2}\!>\!\theta_m |\Phi_{\text{BS}},\Phi_{\rm u}\!\!\right)}_{\mathrm{T}_1}\notag\\
 +&\left(1\!-q\right)\!\times\!\underbrace{\mathbb{P}\left(\frac{\beta H_{n}^{o} R_{n}^{-\eta}}{I_{n}^{\text{intra}}+I^{\text{inter}}+\tilde{\sigma}^2}>\theta_m |\Phi_{\text{BS}},\Phi_{\rm u}\right)}_{\mathrm{T}_2}.
\end{align}
It is obtained by taking the expectation with respect to $b_n$ following Bernoulli ($q$).
We proceed to show the proof of $\mathrm{T}_1$.
\begin{align}
\label{eq:T1}
 \mathrm{T}_1 \overset{(\mathrm{a})}{=}&\mathbb{I}\left({u_{1,m}(\beta)} >0\right)\mathbb{E}\left[e^{-{u_{1,m}(\beta)}^{-1}R_{n}^{\eta}\theta_m \left(I_{n}^{\text{intra}}+I^{\text{inter}}+\tilde{\sigma}^2\right)}\right]\notag\\
 \overset{(\mathrm{b})}{=}&\mathbb{I}\left({u_{1,m}(\beta)} >0\right)e^{- {u_{1,m}(\beta)}^{-1}R_{n}^{\eta}\theta_m\tilde{\sigma}^2}\!\mathbb{E}\left[\prod_{i=n+1}^{N}\prod_{x\in\Phi_{\rm I}}\right.\notag\\
& \left.\!e^{\!-\!{u_{1,m}(\beta)}^{-1}R_{n}^{\eta}\theta_m H_{i}^{o} R_{i}^{-\eta}}\times e^{-{u_{1,m}(\beta)}^{-1}R_{n}^{\eta}\theta_m H_{x}D_x^{-\eta}}\right]\notag\\
 \overset{(\mathrm{c})}{=}&\mathbb{I}\left({u_{1,m}(\beta)} >0\right)e^{- {u_{1,m}(\beta)}^{-1}R_{n}^{\eta}\theta_m\tilde{\sigma}^2}\!\!\prod_{i=n+1}^{N}\prod_{x\in\Phi_{\rm I}}\notag\\
&\frac{{u_{1,m}(\beta)}}{{u_{1,m}(\beta)}\!+\!\theta_m R_{n}^{\eta} D_x^{\!-\!\eta}}\times \frac{{u_{1,m}(\beta)}}{{u_{1,m}(\beta)}\!+\!\theta_m R_{n}^{\eta} R_{i}^{\!-\!\eta}},
\end{align}
where (a) follows from the unit exponential distribution for $H_{n}^{o}$ and lets ${u_{1,m}(\beta)} =(1+\theta_m)\beta-\theta_m$, (b) substitutes the concrete expressions for interferences into the equation. (c) is derived by the unit exponential distribution for $H_{i}^{o}$ and $H_x$ under the assumption that they are i.i.d..
Additionally, the term $\mathrm{T}_2$ of \eqref{eq:gamma1} is derived as
\begin{align}
\label{eq:T2}
&\mathbb{P}\left(\frac{\beta H_{n}^{o} R_{n}^{-\eta}}{I_{n}^{\text{intra}}+I^{\text{inter}}+\tilde{\sigma}^2}>\theta_m |\Phi_{\text{BS}},\Phi_{\rm u}\right)\notag\\
 \overset{(\mathrm{a})}{=}&\mathbb{E}\left[e^{-\beta^{-1}\theta_m R_{n}^{\eta}\left(I_{n}^{\text{intra}}+I^{\text{inter}}+\tilde{\sigma}^2\right)}\right]\notag\\
 \overset{(\mathrm{b})}{=}&e^{\!-\!\beta^{\!-\!1}\theta_m R_{n}^{\eta}\tilde{\sigma}^2}\mathbb{E}\left[\prod_{i=n+1}^{N}\prod_{x\in\Phi_{\rm I}} e^{\!-\!\beta^{\!-\!1}\theta_m R_{n}^{\eta}\left(H_{i}^{o}R_{i}^{\!-\!\eta}+H_x D_x^{-\eta}\right)}\right]\notag\\
 \overset{(\mathrm{c})}{=}&e^{\!-\!\beta^{\!-\!1}\theta_m R_{n}^{\eta}\tilde{\sigma}^2}\!\!\!\! \prod_{i=n+1}^{N}\!\frac{\beta}{\beta\!+\!\theta_m R_{n}^{\eta}R_{i}^{-\eta}}\!\!\prod_{x\in\Phi_{\rm I}}\frac{\beta}{\beta\!+\!\theta_m R_{n}^{\eta}D_x^{-\eta}},
\end{align}
where (a) follows from $H_{n}^{o}\sim \exp (1)$, (b) is derived by substituting the specific expression for intra and inter-cell interference into the equation. (c) takes the expectation of $H_{i}^{o}$ and $H_x$ under the assumption that they are i.i.d. random variables following $\exp(1)$ distribution. The proof of $\mathbb{P}(\gamma_{n2}>\theta_m|\Phi_{\text{BS}},\Phi_{\rm u})$ is analogous to the above, so we omit it here. Thus, the proof of Lemma $1$ is complete.

\subsection{Proof of Theorem $1$}
\label{app:ARR_RSMA}
The average received rate of the typical BS from the $n$-th ranked UE in the RMSA transmission scheme equals to the first moment of the corresponding CRR, which is given by
\begin{align}
M_1^{(n,\text{RS})}\!=\!\mathbb{E}\left[\bar{r}_n^{\text{RS}}\left(\Phi_{\text{BS}},\Phi_{\rm u}\right)\right]
\!\!\overset{(\mathrm{a})}{=}\!\sum_{m=1}^{M}\!\Delta r_m \!\sum_{j=1}^{4}\!\mathbb{E}\left[{I_{j,m,n} \!\left(\beta\right)}\right].
\end{align}
(a) is derived by substituting \eqref{eq:CRR_RSMA}. In the following derivations, we will apply the concrete expressions for {$I_{j,m,n}(\beta)$}. Specifically, we have
\begin{align}
&\mathbb{E}\left[{I_{j,m,n}(\beta)}\right]=\mathbb{E}\left[{A_{j,m,n} \left(\beta\right)f_{j,m,n} \left(\beta\right)}\right]\notag\\
&=\mathbb{E}_{R_{n}}\left[{A_{j,m,n}\left(\beta\right)}\mathbb{E}_{R_{i},D_x}\left[{f_{j,m,n} \left(\beta\right)}\right]\right]\notag\\
&=\int_{0}^{\infty}f_{R_{n}}(r){A_{j,m,n}(\beta)}\mathbb{E}_{R_{i},D_x}\left[{f_{j,m,n}\left(\beta\right)}\right]\mathrm{d}r.
\end{align}
First of all, we take the expectation of ${f_{j,m,n}(\beta)}$ with respect to $R_{i}$ and $D_x$, conditioned on $R_n=r$, i.e.,
\begin{align}
&\mathbb{E}_{R_{i},D_x}\left[{f_{j,m,n}\left(\beta\right)}\right]\!=\!\underbrace{\mathbb{E}_{R_{i}}\!\!\left[\prod_{i=n+1}^{N}\!\!\frac{{u_{j,m}(\beta)}}{{u_{j,m}(\beta)}\!+\!\theta_m \left(r/R_{i}\right)^{\eta}}\right]}_{\mathrm{P}_1}\notag\\
& \times\underbrace{\mathbb{E}_{D_x}\left[\prod_{x\in\Phi_{\rm I}}\frac{{u_{j,m}(\beta)}}{{u_{j,m}(\beta)}+\theta_m \left(r/D_x\right)^{\eta}}\right]}_{\mathrm{P}_2}.
\label{eq:P12}
\end{align}
This equation follows from the independence between intra- and inter-interfering UE distributions. The term $\mathrm{P}_1$ can be further derived as
\begin{align}
\label{eq:P1}
    \mathrm{P}_{1}&\overset{\rm (a)}{=}\frac{(N-n)!}{[1-F(r)]^{N-n}}\int_{\mathcal{A}}\prod_{i=n+1}^{N}\frac{{u_{j,m}(\beta)}  f(r_i)}{{u_{j,m}(\beta)} +\theta_m (r/r_i)^{\eta}}\mathrm{d}\mathbf{r}_n,\notag\\
    &\overset{\rm (b)}{=}2 B_1 \pi \lambda {u_{j,m}(\beta)} (N-n)! e^{(N-n)B_1 \lambda \pi r^2}\notag\\
    &\times \int_{\mathcal{A}}\prod_{i=n+1}^{N} \frac{r_i e^{-B_1 \lambda \pi r_r^2}}{{u_{j,m}(\beta)}  +\theta_m (r/r_i)^{\eta}}\mathrm{d}\mathbf{r}_n
\end{align}
(a) follows the conditional joint PDF in \eqref{eq:cf_r}, where $\mathcal{A}$ denotes the integral area constrained with $r\leq r_{n+1}\leq \cdots \leq r_N$, and $\mathrm{d}\mathbf{r}_n=\mathrm{d}r_{n+1}\cdots \mathrm{d}r_{N}$. (b) is obtained by substituting \eqref{eq:fr}. In addition, the term $\mathrm{P}_2$ in \eqref{eq:P12} can be simplified as follows
\begin{align}
\label{eq:P2}
\mathrm{P}_2
&\overset{(\mathrm{a})}{=}\mathbb{E}_{D_x}\left[\prod_{x\in {\Phi_{\rm A}^{\rm p}}}\left(\frac{{u_{j,m}(\beta)}}{{u_{j,m}(\beta)}+\theta_m \left(r/D_x\right)^{\eta}}\right)^{N}\right]\notag\\
&\overset{\mathrm{(b)}}{=}\exp\left(\!-\!2\pi\lambda\!\int_{0}^{\infty}\!\left[1-\!\left(\frac{{u_{j,m}(\beta)}}{{u_{j,m}(\beta)}+\theta_m \left(r/x\right)^{\eta}}\right)^{N}\right]\right.\notag\\
&\left.\times\left(1-e^{-B_2 \lambda \pi x^2}\right)x\mathrm{d}x\right),
\end{align}
where (a) comes from the fact that $\Phi_{\rm I}$ can be approximated as a Poisson Cluster Process with its parent points {$\Phi_{\rm A}^{\rm p}$, i.e., model A in the main text} and (b) is derived by the probability generating functional (PGFL) of PPP.

\subsection{Proof of Theorem $2$}
\label{Pr:AAR}
Starting from the given equation, that is
\begin{align}
{\bar C_n^{\text{RS}}}=
\underbrace{\mathbb{E}\left[\ln\left(1+\text{SINR}_{n,1}^{\text{RS}}\right)\right]}_{\mathrm{K}_1}+\underbrace{\mathbb{E}\left[\ln\left(1+\text{SINR}_{n,2}^{\text{RS}}\right)\right]}_{\mathrm{K}_2}.
\end{align}
Due to the similarities of $\mathrm{K}_1$ and $\mathrm{K}_2$, we primarily focus on the proof of $\mathrm{K}_1$ here.
\begin{align}
\mathrm{K}_1&={\int_{0}^{\infty}\mathrm{E}\left[\ln \left(1+\text{SINR}_{n,1}^{\text{RS}}\right)\right]f_{R_n}(r)\mathrm{d}r}\notag\\
&=\int_{0}^{\infty}f_{R_n}(r)\int_{0}^{\infty}\mathbb{P}\left(\ln\left(1+\text{SINR}_{n,1}^{\text{RS}}\right)>x\right)\mathrm{d}x\mathrm{d}r\notag\\
&=\int_{0}^{\infty}\!f_{R_n}(r)\int_{0}^{\infty}\! \left(q\!\times\!\mathrm{J}_1+(1\!-\!q)\!\times\!\mathrm{J}_2\right)\mathrm{d}x\mathrm{d}r,
\end{align}
where the expressions for $\mathrm{J}_1$ and $\mathrm{J}_2$ are given by
\begin{align}
\mathrm{J}_1&=\mathbb{P}\left(\frac{\beta H_{n}^{o} R_n^{-\eta}}{H_{n}^{o}R_n^{-\eta}(\!1\!-\!\beta\!)\!+\!I_n^{\text{intra}}\!+\!I_{\text{inter}}\!+\!\tilde{\sigma}^2}\!>\!e^{x}\!-\!1{|R_n=r}\right)\notag\\
\mathrm{J}_2&=\mathbb{P}\left(\frac{\beta H_{n}^{o} R_n^{-\eta}}{I_n^{\text{intra}}+I^{\text{inter}}+\tilde{\sigma}^2}>e^x-1 {|R_n=r}\right).
\end{align}
Similar to the derivation of $\mathrm{T_1}$ and $\mathrm{T}_2$ in \eqref{eq:T1} and \eqref{eq:T2}, we have
\begin{align}
\mathrm{J}_1&=\mathbb{I}\left({L_1(x,\beta)}>0\right)e^{-{L_1(x,\beta)}^{-1}\tilde{\sigma}^2\left(e^x-1\right){r}^{\eta}}\notag\\
&\times \mathbb{E}_{R_i}\left[\prod_{i=n+1}^{N}\frac{{L_1(x,\beta)}}{{L_1(x,\beta)}+(e^x-1)\left({r} / R_i\right)^{\eta}}\right]\notag\\
&\times \mathbb{E}_{D_x}\left[\prod_{x\in\Phi_{\rm I}}\frac{{L_1(x,\beta)}}{{L_1(x,\beta)}+(e^x-1)\left({r} / D_x\right)^{\eta}}\right],
\end{align}
Note that $\mathrm{J}_2$ is obtained from $\mathrm{J}_1$ by replacing ${L_1(x,\beta)}$ with $\beta$. Using the same steps as in \eqref{eq:P1} and \eqref{eq:P2}, and substituting $f_{R_{n}}(r)$, we obtain the result and the proof is complete.

\subsection{Proof of Corollary $2$}
\label{Pr:Co4}
Consider the definition of moments, we give the $b$-th moments of the CRR as follows
\begin{align}
&\mathbb{E}\left[ \left({\bar{r}_n^{\text{NO}}(\Phi_{\text{BS}},\Phi_{\rm u})}\right)^{b}\right]
\! \overset{(\mathrm{a})}{=}\!\!\!\sum_{C_1,C_2}\!\!\!A_b\! \left(\prod_{m=1}^{M}\!\Delta r_{m}^{n_{m}}\!\right)\!\mathbb{E}\!\left[\prod_{m=1}^{M}\!{e^{\!-\!\theta_m R_{n}^{\eta}\tilde{\sigma}^{2}n_m }}\right.\notag\\
&\left.\times\left(\prod_{i=n+1}^{N}\!\frac{1}{(\theta_m R_{n}^{\eta}R_{i}^{-\eta}\!+\!1)^{n_m}}\!\times\!\prod_{x\in\Phi_{\rm I}}\!\frac{1}{(\theta_m R_{n}^{\eta}D_{x}^{-\eta}\!+\!1)^{n_m}}\right)\right]\notag\\
&\!=\!\!\sum_{C_1,C_2}\!\!A_b \!\left(\prod_{m=1}^{M}\!\Delta r_{m}^{n_{m}}\right)\mathbb{E}_{R_{n}}\!\left[{e^{\!-\!\theta_m R_{n}^{\eta}\tilde{\sigma}^2 \Omega}}\mathbb{E}_{R_{i}}\!\!\left[\prod_{i=n+1}^{N}\!\prod_{m=1}^{M}\right.\right.\notag\\
&\left.\left.\frac{1}{(\theta_m R_{n}^{\eta}R_{i}^{-\eta}\!+\!1)^{n_m}}\right]\mathbb{E}_{D_x}\left[\prod_{x\in\Phi_{\rm I}}\prod_{m=1}^{M}\frac{1}{(\theta_m R_{n}^{\eta}D_{x}^{-\eta}\!+\!1)^{n_m}}\right]\right],\label{equ:29}
\end{align}
where $\Omega=\sum_{m=1}^{M} \theta_m n_m$, (a) follows from multi-nominal series and the moment generation function of $\bar{r}(\Phi_{\text{BS}},\Phi_{\rm u})$ with $C_1:n_1,n_2,\!\cdots,\!n_M\geq 0$. $C_2:n_1\!+\!n_2\!+\!\cdots\!+\!n_M\!=\!b$, and $A_b = \frac{b!}{n_{1}! n_{2}!\cdots n_{M}!}$.
Inside \eqref{equ:29}, the derivations of taking the expectations with respect to $R_i$ and $D_x$ is similar to that of $\mathrm{P}_1$ and $\mathrm{P}_2$ in \eqref{eq:P1} and \eqref{eq:P2}, respectively. For brevity, we omit the related process here. Thus, the proof is complete.

%% file: reference.bib
@article{wang2023road,
  title={On the road to {6G}: Visions, requirements, key technologies, and testbeds},
  author={Wang, Cheng-Xiang and You, Xiaohu and Gao, Xiqi and Zhu, Xiuming and Li, Zixin and Zhang, Chuan and Wang, Haiming and Huang, Yongming and Chen, Yunfei and Haas, Harald and others},
  journal={IEEE Commun. Surveys Tuts.},
  volume={25},
  number={2},
  pages={905--974},
  month={2nd Quart.},
  year={2023},
  publisher={IEEE}
}

@article{haenggi2017user,
  title={User point processes in cellular networks},
  author={Haenggi, Martin},
  journal={IEEE Wireless Commun. Lett.},
  volume={6},
  number={2},
  pages={258--261},
  month={Feb.},
  year={2017},
  publisher={IEEE}
}

@article{wang2017meta,
  title={The meta distribution of the {SIR} for cellular networks with power control},
  author={Wang, Yuanjie and Haenggi, Martin and Tan, Zhenhui},
  journal={IEEE Trans. Commun.},
  volume={66},
  number={4},
  pages={1745--1757},
  month={Dec.},
  year={2017},
  publisher={IEEE}
}

@article{haenggi2015meta,
  title={The meta distribution of the {SIR} in {P}oisson bipolar and cellular networks},
  author={{Haenggi, Martin}},
  journal={IEEE Trans. Wireless Commun.},
  volume={15},
  number={4},
  pages={2577--2589},
  month={Dec.},
  year={2015},
  publisher={IEEE}
}

@inproceedings{martin2017interference,
  title={Interference-aware muting for the uplink of heterogeneous cellular networks: A stochastic geometry approach},
  author={Martin-Vega, Francisco Javier and Aguayo-Torres, M Carmen and G{\'o}mez, Gerardo and Di Renzo, Marco},
  booktitle={Proc. IEEE Int. Conf. Commun. (ICC)},
  pages={1--6},
  month={May},
  year={2017},
  address = {Paris, France}
}

@book{haenggi2012stochastic,
  title={{S}tochastic {G}eometry for {W}ireless {N}etworks},
  author={Haenggi, Martin},
  year={2012},
  publisher={Cambridge University Press}
}

@article{salehi2018meta,
  title={Meta distribution of {SIR} in large-scale uplink and downlink {NOMA} networks},
  author={Salehi, Mohammad and Tabassum, Hina and Hossain, Ekram},
  journal={IEEE Trans. Commun.},
  volume={67},
  number={4},
  pages={3009--3025},
  month={Dec.},
  year={2018},
  publisher={IEEE}
}

@inproceedings{guo2025stochastic,
  title={Stochastic Geometry-Based {MCS} Adaption Analysis for Uplink Cellular Networks},
  author={Guo, Xinyi and Liu, Qiong and Wang, Shanshan and You, Li},
  booktitle={Proc. IEEE Wireless Commun. Netw. Conf. (WCNC)},
  pages={1--6},
  month={Mar.},
  year={2025},
  address={Milan, Italy}
}

@article{tabassum2017modeling,
  title={Modeling and analysis of uplink non-orthogonal multiple access in large-scale cellular networks using {p}oisson cluster processes},
  author={Tabassum, Hina and Hossain, Ekram and Hossain, Jahangir},
  journal={IEEE Trans. Commun.},
  volume={65},
  number={8},
  pages={3555--3570},
  month={Apr},
  year={2017},
  publisher={IEEE}
}

@article{lu2021stochastic,
  title={Stochastic geometry analysis of spatial-temporal performance in wireless networks: A tutorial},
  author={Lu, Xiao and Salehi, Mohammad and Haenggi, Martin and Hossain, Ekram and Jiang, Hai},
  journal={IEEE Commun. Surv. Tuts.},
  volume={23},
  number={4},
  pages={2753--2801},
  month={4th Quart.},
  year={2021},
  publisher={IEEE}
}

@article{wang2019meta,
  title={On the meta distribution in spatially correlated non-{P}oisson cellular networks},
  author={Wang, Shanshan and Di Renzo, Marco},
  journal={EURASIP J. Wireless Commun. Netw.},
  volume={2019},
  pages={1--11},
  month={Jul.},
  year={2019},
  publisher={Springer}
}

@article{clerckx2023primer,
  title={A primer on rate-splitting multiple access: {T}utorial, myths, and frequently asked questions},
  author={Clerckx, Bruno and Mao, Yijie and Jorswieck, Eduard A and Yuan, Jinhong and Love, David J and Erkip, Elza and Niyato, Dusit},
  journal={IEEE J. Sel. Areas Commun.},
  volume={41},
  number={5},
  pages={1265--1308},
  month={Feb.},
  year={2023},
  publisher={IEEE}
}

@article{rimoldi1996rate,
  title={A rate-splitting approach to the {G}aussian multiple-access channel},
  author={Rimoldi, Bixio and Urbanke, R{\"u}diger},
  journal={IEEE Trans. Inf. Theory},
  volume={42},
  number={2},
  pages={364--375},
  month={Mar.},
  year={1996},
  publisher={IEEE}
}

@article{yang2020sum,
  title={Sum-rate maximization of uplink rate splitting multiple access ({RSMA}) communication},
  author={Yang, Zhaohui and Chen, Mingzhe and Saad, Walid and Xu, Wei and Shikh-Bahaei, Mohammad},
  journal={IEEE Trans. Mob. Comput.},
  volume={21},
  number={7},
  pages={2596--2609},
  month={Nov.},
  year={2020},
  publisher={IEEE}
}

@article{liu2020rate,
  title={Rate splitting for uplink {NOMA} with enhanced fairness and outage performance},
  author={Liu, Hongwu and Tsiftsis, Theodoros A and Kim, Kyeong Jin and Kwak, Kyung Sup and Poor, H Vincent},
  journal={IEEE Trans. Wireless Commun.},
  volume={19},
  number={7},
  pages={4657--4670},
  month={Apr.},
  year={2020},
  publisher={IEEE}
}

@inproceedings{ha2020coordinated,
  title={Coordinated rate splitting multiple access for multi-cell downlink networks},
  author={Ha, Nohgyeom and Shin, Wonjae and Vaezi, Mojtaba and Poor, H Vincent},
  booktitle={Proc. Asilomar Conf. Signals Syst. Comput.},
  pages={996--1001},
  month={Nov.},
  year={2020},
  address={Pacific Grove, CA, USA}
}

@inproceedings{mao2019rate,
  title={Rate-splitting multiple access for coordinated multi-point joint transmission},
  author={Mao, Yijie and Clerckx, Bruno and Li, Victor OK},
  booktitle={Proc. IEEE Int. Conf. Commun. Workshops (ICC Workshops)},
  pages={1--6},
  month={May},
  year={2019},
  address={Shanghai, China}
}

@article{zhu2023rate,
  title={Rate-splitting multiple access in multi-cell dense networks: A stochastic geometry approach},
  author={Zhu, Qiao and Qian, Zhihong and Clerckx, Bruno and Wang, Xue},
  journal={IEEE Trans. Veh. Technol.},
  volume={72},
  number={12},
  pages={15844--15857},
  month={Jul.},
  year={2023},
  publisher={IEEE}
}

@inproceedings{dizdar2020rate,
  title={Rate-splitting multiple access for downlink multi-antenna communications: Physical layer design and link-level simulations},
  author={Dizdar, Onur and Mao, Yijie and Han, Wei and Clerckx, Bruno},
  booktitle={Proc. IEEE Int. Symp. Person. Indoor Mobile Radio Commun. (PIMRC)},
  pages={1--6},
  month={Aug.},
  year={2020},
  address={London, UK}
}

@article{mishra2021rate,
  title={Rate-splitting multiple access for downlink multiuser {MIMO}: Precoder optimization and PHY-layer design},
  author={Mishra, Anup and Mao, Yijie and Dizdar, Onur and Clerckx, Bruno},
  journal={IEEE Trans. Commun.},
  volume={70},
  number={2},
  pages={874--890},
  month={Dec.},
  year={2021},
  publisher={IEEE}
}

@article{ding2024next,
  title={Next generation multiple access for {IMT} towards 2030 and beyond},
  author={Ding, Zhiguo and Schober, Robert and Fan, Pingzhi and Poor, H Vincent},
  journal={Sci. China Inf. Sci.},
  volume={67},
  number={6},
  pages={166301},
  month={Jun.},
  year={2024},
  publisher={Springer}
}

@article{andrews2011tractable,
  title={A tractable approach to coverage and rate in cellular networks},
  author={Andrews, Jeffrey G and Baccelli, Fran{\c{c}}ois and Ganti, Radha Krishna},
  journal={IEEE Trans. Commun.},
  volume={59},
  number={11},
  pages={3122--3134},
  month={Oct.},
  year={2011},
  publisher={IEEE}
}

@book{david2004order,
  title={Order {S}tatistics},
  author={David, Herbert A and Nagaraja, Haikady N},
  year={2004},
  publisher={John Wiley \& Sons}
}

@article{mao2022rate,
  title={Rate-splitting multiple access: Fundamentals, survey, and future research trends},
  author={Mao, Yijie and Dizdar, Onur and Clerckx, Bruno and Schober, Robert and Popovski, Petar and Poor, H Vincent},
  journal={IEEE Commun. Surveys \& Tuts.},
  volume={24},
  number={4},
  pages={2073--2126},
  month={4th Quart.},
  year={2022},
  publisher={IEEE}
}

@article{6773024,
  title={A mathematical theory of communication},
  author={Shannon, C. E.},
  journal={The Bell System Technical Journal}, 
  year={1948},
  volume={27},
  number={3},
  pages={379-423},
  month={Jul.}
}

@article{liang2019non,
  title={Non-orthogonal multiple access ({NOMA}) in uplink {P}oisson cellular networks with power control},
  author={Liang, Yanan and Li, Xu and Haenggi, Martin},
  journal={IEEE Trans. Commun.},
  volume={67},
  number={11},
  pages={8021--8036},
  year={2019},
  month={Aug.},
  publisher={IEEE}
}

@article{tegos2025distributed,
  title={Distributed uplink rate splitting multiple access ({DU-RSMA}): Principles and performance analysis},
  author={Tegos, Apostolos A and Xiao, Yue and Tegos, Sotiris A and Karagiannidis, George K and Diamantoulakis, Panagiotis D},
  journal={IEEE Trans. Wireless Commun.},
  pages={1--1},
  year={2025},
  month={Jul.},
  publisher={IEEE}
}

@article{katwe2022rate,
  title={Rate splitting multiple access for sum-rate maximization in {IRS} aided uplink communications},
  author={Katwe, Mayur and Singh, Keshav and Clerckx, Bruno and Li, Chih-Peng},
  journal={IEEE Trans. Wireless Commun.},
  volume={22},
  number={4},
  pages={2246--2261},
  month={Oct.},
  year={2022},
  publisher={IEEE}
}

@article{yang2021optimization,
  title={Optimization of rate allocation and power control for rate splitting multiple access {(RSMA)}},
  author={Yang, Zhaohui and Chen, Mingzhe and Saad, Walid and Shikh-Bahaei, Mohammad},
  journal={IEEE Trans. Commun.},
  volume={69},
  number={9},
  pages={5988--6002},
  month={Jun.},
  year={2021},
  publisher={IEEE}
}

@article{chen2020performance,
  title={On the performance of cluster-based {MIMO-NOMA} in multi-cell dense networks},
  author={Chen, Guangji and Qiu, Ling and Ren, Chenhao},
  journal={IEEE Trans. Commun.},
  volume={68},
  number={8},
  pages={4773--4787},
  month={Apr.},
  year={2020},
  publisher={IEEE}
}

@article{kusaladharma2019outage,
  title={Outage performance and average rate for large-scale millimeter-wave {NOMA} networks},
  author={Kusaladharma, Sachitha and Zhu, Wei-Ping and Ajib, Wessam},
  journal={IEEE Trans. Wireless Commun.},
  volume={19},
  number={2},
  pages={1280--1291},
  month={Nov.},
  year={2019},
  publisher={IEEE}
}

@article{yang2024joint,
  title={Joint inter-group common stream superposition and user grouping-based {RSMA} in overloaded systems},
  author={Yang, Junliang and Gao, Shaoshuai and Lu, Yan and Yang, Zhao and Tu, Guofang},
  journal={IEEE Wireless Commun. Lett.},
  volume={13},
  number={9},
  pages={2497--2501},
  month={Jul.},
  year={2024},
  publisher={IEEE}
}

@article{lyu2024rate,
  title={Rate-splitting multiple access for overloaded multi-group multicast: A first experimental study},
  author={Lyu, Xinze and Aditya, Sundar and Clerckx, Bruno},
  journal={IEEE Trans. Broadcast.},
  volume={71},
  number={1},
  pages={30--41},
  month={Mar.},
  year={2024},
  publisher={IEEE}
}

@article{vu2025outage,
  title={Outage, capacity, and error performance of downlink {RSMA}-based systems: Analysis and resource optimization},
  author={Vu, Thai-Hoc and Da Costa, Daniel Benevides and Kim, Sunghwan and Pham, Quoc-Viet},
  journal={IEEE Trans. Commun.},
  volume={73},
  number={8},
  pages={6868--6883},
  month={Feb.},
  year={2025},
  publisher={IEEE}
}

@article{krishnamoorthy2022downlink,
  title={Downlink {MIMO-RSMA} with successive null-space precoding},
  author={Krishnamoorthy, Aravindh and Schober, Robert},
  journal={IEEE Trans. Wireless Commun.},
  volume={21},
  number={11},
  pages={9170--9185},
  month={May},
  year={2022},
  publisher={IEEE}
}

@article{zamani2020optimizing,
  title={Optimizing weighted-sum energy efficiency in downlink and uplink {NOMA} systems},
  author={Zamani, Mohammad Reza and Eslami, Mohsen and Khorramizadeh, Mostafa and Zamani, Hojatollah and Ding, Zhiguo},
  journal={IEEE Trans. Veh. Technol.},
  volume={69},
  number={10},
  pages={11112--11127},
  month={Jul.},
  year={2020},
  publisher={IEEE}
}

@article{jiang2023rate,
  title={Rate-splitting multiple access for uplink massive {MIMO} with electromagnetic exposure constraints},
  author={Jiang, Hanyu and You, Li and Elzanaty, Ahmed and Wang, Jue and Wang, Wenjin and Gao, Xiqi and Alouini, Mohamed-Slim},
  journal={IEEE J. Sel. Areas. Commun.},
  volume={41},
  number={5},
  pages={1383--1397},
  month={Jan.},
  year={2023},
  publisher={IEEE}
}
